\documentclass{aa}  

\usepackage{graphicx}
\usepackage{subcaption}
\usepackage{txfonts}
\usepackage{textcomp} 
\usepackage{siunitx}  
\DeclareSIUnit\gauss{Gs}
\DeclareSIUnit\tecu{TECU}
\DeclareSIUnit\jansky{Jy}
\DeclareSIUnit\beam{beam}
\DeclareSIUnit\rad{rad}
\usepackage{pifont} 
\newcommand{\cmark}{\ding{51}}%
\newcommand{\xmark}{\ding{55}} %
\newcommand*\dif{\mathop{}\!\mathrm{d}}

\newcommand{\TEC}{\ensuremath{\mathit{TEC}}\xspace}
\newcommand{\dTEC}{\ensuremath{\dif\mathit{TEC}}\xspace}
\newcommand{\vTEC}{\ensuremath{\mathit{vTEC}}\xspace}
\newcommand{\sTEC}{\ensuremath{\mathit{sTEC}}\xspace}
\newcommand{\RM}{\ensuremath{\mathit{RM}}\xspace}
\newcommand{\dRM}{\ensuremath{\dif\mathit{RM}}\xspace}
\newcommand{\SEFD}{\ensuremath{\mathit{SEFD}}\xspace}
\newcommand{\rms}{\ensuremath{\mathit{rms}}\xspace}
\usepackage{makecell} 

\usepackage{natbib}
\usepackage{hyperref}

\begin{document}

   \title{Investigating ionospheric calibration for LOFAR 2.0 \\with simulated observations}
   \titlerunning{Investigating LOFAR 2.0 calibration with simulations} 

   \author{H.~W.~Edler\inst{1}
          \and
          F.~de~Gasperin\inst{1,2}
          \and
          D.~Rafferty\inst{1}
          }

   \institute{Hamburger Sternwarte, University of Hamburg,
              Gojenbergsweg 112, D-21029, Hamburg, Germany\\
              \email{henrik.edler@hs.uni-hamburg.de}
              \and
              INAF - Istituto di Radioastronomia,
              via P. Gobetti 101, 40129, Bologna, Italy}

   \date{Preprint May 10, 2021}

  \abstract
   {A number of hardware upgrades for the Low-Frequency Array (LOFAR) are currently under development. These upgrades are collectively referred to as the LOFAR\,2.0 upgrade. The first stage of LOFAR\,2.0 will introduce a distributed clock signal and allow for simultaneous observation with all the low-band and high-band antennas of the array.}
   {Our aim is to provide a tool for accurate simulations of LOFAR\,2.0.}
   {We present a software to simulate LOFAR and LOFAR\,2.0 observations, which includes realistic models for all important systematic effects such as the first and second order ionospheric corruptions, time-variable primary-beam attenuation, station based delays and bandpass response. The ionosphere is represented as a thin layer of frozen turbulence. Furthermore, thermal noise can be added to the simulation at the expected level.
   We  simulate a full 8-hour simultaneous low- and high-band antenna observation of a calibrator source and a target field with the LOFAR\,2.0 instrument. The simulated data is calibrated using readjusted LOFAR calibration strategies. We examine novel approaches of solution-transfer and joint calibration to improve direction-dependent ionospheric calibration for LOFAR.}
   {We find that the calibration of the simulated data behaves very similarly to a real observation and reproduces characteristic properties of LOFAR data such as realistic solutions and image quality. We analyze strategies for direction-dependent calibration of LOFAR\,2.0 and find that the ionospheric parameters can be determined most accurately when combining the information of the high-band and low-band in a joint calibration approach. In contrast, the transfer of total electron content solutions from the high-band to the low-band shows good convergence but is highly susceptible to the presence of non-ionospheric phase errors in the data.}
   {}


   \maketitle
%

\section{Introduction}

The Low-Frequency Array (LOFAR) is an interferometer working at low (<\SI{1}{\giga\hertz}) and ultra-low (<\SI{100}{\mega\hertz}) radio frequencies. Most of its stations are located in the Netherlands and it is operational since 2010 \citep{vanHaarlem2013}.
It observes in a spectral band close to the low-frequency edge of the radio window, which received reinforced attention in the last decade.
Together with other new generation instruments, such as the Murchison Widefield Array (MWA, \cite{mwa2013}) and the Long Wavelength Array (LWA, \cite{lwa2012}), LOFAR pioneers the technological ground for the ambitious plans on the Square Kilometer Array (SKA, \cite{ska2009}), which include the low-frequency interferometer SKA-low.

Among the most important research prospects in low-frequency radio astronomy are deep surveys, in which phenomena such as the extended emission in galaxy clusters and high-redshift radio galaxies are studied. These sources oftentimes remain undetectable at higher frequencies due to their steep spectral energy distribution. 
Further science cases include the search for the epoch of reionization hydrogen line, cosmic magnetism, radio detection of exoplanets, mapping of the star-formation history of the universe as well as the observation of pulsars and radio transients.

A challenging limitation for the scientific value of wide-field images is imposed by the direction-dependent corruptions the ionosphere introduces for earthbound measurements of cosmic radio signals. This is especially problematic at ultra-low frequencies, where the impact of the ionosphere becomes increasingly dramatic.  
To accurately account for ionospheric effects during calibration, the development of a new generation of direction-dependent calibration strategies together with the corresponding software environment was initiated during the last decade \citep{Tasse2021,deGasperin2020, Albert2020,vanWeeren2016}. 
Nevertheless, excellent image fidelity is still hard to achieve with LOFAR for complex fields (e.g. towards the Galactic plane), during exceptionally active ionospheric conditions, and for the lowest part of the frequency band ($\lesssim\SI{40}{\mega\hertz}$). 

This motivates the LOFAR\,2.0 upgrade, which is going to increase system performance, especially for the low-band antenna (LBA) part of the array (Hessels et al. in prep.). This upgrade will allow for simultaneous observations with the low-band and high-band antennas (HBA), and thus enable novel calibration algorithms which combine information of both antennas.

To make efficient use of the upgraded hardware, it is important to examine new calibration strategies for LOFAR\,2.0 prior to the deployment of the upgraded hardware. We developed the \emph{LOFAR Simulation Tool} (\texttt{LoSiTo}) to simulate a full 8-h simultaneous calibrator and target field observation with the upgraded LOFAR 2.0 system. First, we calibrate the simulated data for instrumental and direction-independent effects, relying mostly on standard LOFAR tools. Then, we investigate novel approaches for direction-dependent calibration with the aim of improving the ionospheric calibration of the LBA system. 

This paper is arranged as follows: in \autoref{sec:lofar2.0}, we discuss the current limitations of LOFAR and the improvements which will be introduced by LOFAR\,2.0. In \autoref{sec:corruptions}, we present models for the instrumental and ionospheric systematic errors in LOFAR\,2.0 observations and estimate the noise level of the upgraded array. The implementation of these models in a simulation software as well as the setup of the simulated observation is described in \autoref{sec:losito}. Next in \autoref{sec:calibration}, we proceed with the calibration of the simulated data and investigate how the direction-dependent calibration for LBA could be improved by solution-transfer or joint calibration.

\section{LOFAR and Future Upgrades}

\label{sec:lofar2.0}

LOFAR is an array consisting of 55 stations spread across Europe. Three different station types with different hardware configurations exist: the central part of the array is composed of 24 core stations (CS), these are spread across an area in the Northeast of the Netherlands which is less than \SI{4}{\kilo\meter} wide. Other 16 stations are also located in the Netherlands, further away from the core, they are called remote stations (RS). The international part of the array consists of 15 more stations in eight European countries. Each LOFAR station hosts a field of dipoles operating as phased array, two drastically different dipole designs are employed. The LBA are sensitive in the frequency range of \SI{10}{\mega\hertz} -- \SI{90}{\mega\hertz}. Additionally, the upper part of the frequency bandwidth of LOFAR is covered by the  HBA, which are sensitive from \SI{110}{\mega\hertz} to \SI{240}{\mega\hertz}. Each Dutch station is equipped with 96 LBA dipoles and 2$\times$24 (CS) respectively 48 (RS) HBA tiles. 
The antenna signals are locally digitized in receiver units (RCUs), there are a total of 48 RCUs per Dutch station. Hence, operation is limited to either all 48 HBA tiles or 48 out of the 96 LBA dipoles \citep{vanHaarlem2013}. Consequently, in LBA not all the existing dipoles are utilized and the potential sensitivity is not reached.

Currently, the clock signals for all remote and international stations are provided by GPS-synchronized rubidium clocks. These clocks show the presence of a $\mathcal{O}(\SI{10}{\nano\second\per\hour})$ clock drift which causes non-negligible phase errors in the data \citep{deGasperin2019,vanWeeren2016}. 

The LOFAR\,2.0 upgrade is designed to overcome these limitations and consists of multiple projects (Hessels et al. in prep). In this work, we focus on the changes introduced by the \emph{Digital Upgrade for Premier LOFAR Low-band Observing} (DUPLLO). DUPLLO is a funded project \citep{nwo2018}, its main focus lies on improving the performance of the LBA system.
Among the key components of DUPLLO is the deployment of improved station electronics. The number of RCUs per station will be tripled, this is going to allow for simultaneous observations with all 96 LBA dipoles as well as the 48 HBA tiles. Consequences will be an increase of sensitivity and field of view (FoV) in LBA observations and the possibility of innovative calibration algorithms which can exploit the simultaneous observation to derive more accurate direction-dependent calibration solutions. 
The second improvement introduced by DUPLLO will be a new clock system. A ``White Rabbit'' Ethernet based timing distribution system \citep{Lipinski2011} will be employed to synchronize all Dutch stations with a single clock \citep{Cappellen2019}. 

Roll-out of LOFAR\,2.0 stage 1, which includes DUPLLO, is anticipated for April 2022 and full operation is expected for December 2023 \citep{Cappellen2019}.

\section{Model Corruptions}

\label{sec:corruptions}
To accurately simulate LOFAR observations, realistic models for the corrupting effects present in the observations are required.
Systematic effects related to physical and instrumental phenomena may be extracted from calibration solutions due to their characteristic properties and dependencies. In \citet{deGasperin2019}, measurements of the bright calibrator source \textit{3C196} were analyzed and all major corrupting effects present in LOFAR observations could be isolated from the calibration solutions. We take these effects  as a reference for the systematics that need to be included in realistic simulations. A summary of these effects together with the dominant noise components is presented in \autoref{tab:systematics}.

\begin{table*}
\centering
\caption{Systematic effects and noise components in LOFAR observations.}
\begin{tabular}{llllll}
\thead[l]{Effect} & \thead[l]{Type of \\ Jones matrix} & \thead[l]{Phase or \\Amplitude?} & \thead[l]{Frequency \\ dependence} & \thead[l]{Direction \\ dependent?} & \thead[l]{Time \\ dependent?} \\\hline
Clock drift & scalar & phase & $\propto\nu$ & \xmark & \cmark(minutes) \\
Polarization alignment & diagonal & phase & $\propto\nu$ & \xmark & \xmark \\ 
Ionosphere 1st ord. (dispersive delay) & scalar & phase & $\propto\nu^{-1}$ & \cmark & \cmark(seconds) \\ 
Ionosphere 2nd ord. (Faraday rotation) & rotation & both & $\propto\nu^{-2}$ & \cmark & \cmark(seconds)\\
Ionosphere 3rd ord. & scalar & phase & $\propto\nu^{-3}$ & \cmark & \cmark(seconds) \\
Ionosphere scintillations & diagonal & amplitude & - & \cmark & \cmark(seconds)  \\
Dipole beam & full-Jones & both & complex & \cmark & \cmark(minutes) \\
Array factor & scalar & both & complex & \cmark & \cmark(minutes) \\
Bandpass & diagonal & amplitude & complex & \xmark & \xmark\\
Sky noise & additive & both & $\propto\nu^{-2.57}$\tablefootmark{a} & \cmark & \cmark(random) \\ 
Instrumental noise & additive & both & complex  & \xmark & \cmark(random)  
\end{tabular}
\tablefoot{This table is an extended version of the one in \citet{deGasperin2019} and contains the noise components discussed in \citet{vanHaarlem2013}.\\
\tablefoottext{a}{Sky noise spectral index from \citet{guzman2011} and valid for the North Galactic Pole.}
}
\label{tab:systematics}
\end{table*}

The mathematical framework of radio-interferometry is given by the radio interferometer measurement equation \citep[hereinafter RIME; see][]{hamaker1996,smirnov2011}. The RIME connects the complex visibility $\mathbf{V}$, which is the quantity measured at the interferometer, to the sky brightness distribution $\mathbf{B}$. Systematic effects enter the RIME as \emph{Jones-matrices} $\mathbf{J}_a$. Jones matrices are $2\times2$ matrices defined on the  linear (or equivalently, circular) polarization basis of the electromagnetic field. If multiple effects are present, the total Jones matrix is the matrix product of the individual matrices. The order for matrix-multiplication is given by the order in which the effects occur along the signal path: $\mathbf{J}_{total} = \mathbf{J}_1\cdot...\cdot\mathbf{J}_n$. The shape of the Jones-matrix is determined by the polarization-dependence of the underlying effect. A full formulation of the RIME is given by:
\begin{equation}
    \mathbf{V}_{ab} = \iint \mathbf{J}_a(l,m) \mathbf{B}(l,m)\mathbf{J}_b^\dag(l,m) e^{-2\pi i \left(ul + vm + wn\right)} \frac{\dif l \dif m}{n},
    \label{eq:rime}
\end{equation} 
where $l,m$ and $n=\sqrt{1-l^2-m^2}$ are the components of the source direction unit vector and $u,v,w$ are the components of the baseline vector measured in wavelengths.

\subsection{Ionosphere model}

The ionized plasma of the upper atmosphere interferes with radioastronomic observations in a variety of ways. Series-expansion of the diffractive index $n(\nu)$ in $\nu^{-1}$ allows to describe this interference by a few simple effects \citep{dattabarua2008}. 
The dominant ionospheric effect is a dispersive delay which expresses as a scalar phase error $\Delta\phi$ with a characteristic frequency dependence of $\propto \nu^{-1}$ \citep{mevius2016}:
\begin{equation}\label{eq:dispersive_delay}
    \Delta\phi = -84.48 \left[\frac{\dif\mathit{TEC}}{1\, \mathrm{TECU}}\right]\left[\frac{\SI{100}{\mega\hertz}}{\nu}\right]\,\mathrm{rad}.
\end{equation}
This phase error depends on the line of sight integrated electron density $N_e$, which is referred to as the total electron content (\TEC):
\begin{equation}
    \mathit{TEC} = \int N_e {\dif}l.
\end{equation}
The \TEC is most commonly measured in total electron content units (TECU, 1$\,$TECU = \SI{10e16}{\per\square\meter}).
Since the \emph{RIME} is insensitive to a global scalar phase, only the differential \TEC between the stations is relevant for the dispersive delay. 

Another ionospheric effect that is non-negligible at the frequency range of LOFAR is Faraday rotation. This effect is of second order in $\nu^{-1}$ and hence, especially problematic at the lowest frequencies. It expresses as rotation in the plane of linear polarization.
The line-of-sight contribution to the Faraday rotation depends on the magnetic field $\vec{B}$ and the free electron density and can be summarized into the \textit{rotation measure} (\RM):
\begin{equation}
    \RM =  \frac{e^3}{8\pi^2\epsilon_0 m_e^2 c^3} \int N_e(l)\lVert{\vec B}\rVert \cos(\theta)\dif l. 
    \label{eq:rm}
\end{equation}
Here, $e$ is the electron charge, $m_e$ the electron mass, $c$ the speed of light in vacuum, $\epsilon_0$ the vacuum permittivity and $\theta$ the angle between the magnetic field vector and the line-of-sight. 
The corresponding rotation angle $\beta$ is given by:
\begin{equation}
    \beta = \RM\cdot \left(\frac{c}{\nu}\right)^2.
\end{equation}
Ionospheric rotation does also affect unpolarized signals: the magnetic flux density and the free electron number density vary between signal paths. This introduces a relative rotation angle between different stations and source directions. This effect is known as \textit{differential Faraday rotation} and can, in linear polarization basis, be described by a rotational Jones matrix.  
This differential rotation causes amplitude and phase errors and may de-correlate the signal in extreme cases.
The impact of this effect becomes important for frequencies below \SI{80}{\mega\hertz} and baselines longer than \SI{10}{\kilo\meter} \citep{maaijke2018}.

The next important higher order effect is of third order in $\nu^{-1}$ and manifests as dispersive delay, similar to the first order effect. In \cite{deGasperin2018}, it was found that this effect is only important for frequencies below \SI{40}{\mega\hertz}. Since the effect is negligible for most of the frequency range observed by LOFAR and also quadratic in the free electron density and hence, harder to model compared to the first and second order effects, we have chosen to not include it in the simulation.

To obtain a realistic model of the ionosphere, a number of characteristic properties have to be considered. The scale and structure of the electron content must be in reasonable agreement with reality. Furthermore, the electron distribution must be spatially and temporally coherent.
We employ the thin-layer model to describe the free electron density of the ionosphere. It represents the ionosphere as two-dimensional spherical shell at a height of $h_{ion}$ around the Earth. This approximation is motivated by the vertical structure of the ionosphere. The majority of the free electrons are constrained within the ionospheric $F$-layer between \SI{200}{\kilo\meter} and \SI{450}{\kilo\meter}.
Contracting the three-dimensional structure onto a two-dimensional sphere drastically reduces the complexity of the model while maintaining many of the important characteristics, such as spatial coherency and to some accuracy, the elevation dependence of the projected electron content \citep[see][]{martin2016}. 

In the thin-layer approximation, the ionosphere is fully parameterized by a two-dimensional distribution of the vertical total electron content (\vTEC). This distribution is hereinafter referred to as a \textit{\TEC-screen}.
The \TEC value corresponding to a specific signal path is evaluated at the ionospheric pierce point, which is defined as the point where the source direction vector \emph{pierces} the \TEC-screen.

For sources which are not directly at the zenith, an air-mass factor has to be taken into account to derive the slant \TEC (\sTEC) along the line of sight. Projection leads to an increase of the \TEC for directions further away from the zenith:
\begin{equation}
   \sTEC = \frac{\vTEC}{\cos(\theta_{ion})}, 
\end{equation}
where $\theta_{ion}$ is the pierce angle.
Since we model the Earth as a sphere of radius $R_E$, the pierce angle can be calculated from the source elevation $\theta'$ according to the law of sines \citep{martin2016}:
\begin{equation}
    \sin(\theta_{ion}) = \frac{R_E}{R_E + h_{ion}} \sin(\theta'),
\end{equation}
which allows to derive the corresponding \sTEC for any direction:
\begin{equation}
    \sTEC = \frac{\vTEC}{\sqrt{1 - \frac{R_E }{R_E + h_{ion}}\sin(\theta')}}.
    \label{eq:sTEC}
\end{equation}

A substantial fraction of ionospheric inhomogeneity can be attributed to turbulent phenomena \citep{materassi2019,giannattasio2018}.
Ionospheric turbulence can be approximately described by a process where energy is injected into a system with high Reynolds number at large spacial scales and iteratively distributed to smaller scales \citep{thompson2017}. This self-similarity leads to a refractive-index power spectrum $\Phi(k)$ that follows a power-law shape:
\begin{equation}
    \Phi(k)\propto k_{}^{-\beta},
\end{equation}
where $k$ is the spatial frequency. In case of pure Kolmogorov turbulence, the spectral index $\beta=11/3$ \citep{tatarski1961,tatarski1971}.  
Taking the finite outer scale $L_0$ of the turbulence into consideration yields:
\begin{equation}
    \Phi(k)\propto \left(k^2+L_0^{-2}\right)^{-\beta},
\end{equation}
which in the $\beta=11/3$ case describes the von Kármán spectrum \citep{VonKarman1948}.

Previous studies with LOFAR confirmed the power-law shape, but found a slightly steeper spectrum of $\beta=3.89\pm0.01$ \citep{mevius2016,deGasperin2019}\footnote{Derived from the phase structure function spectral index $\alpha$ as $\beta = \alpha - 2$ according to \citet{Boreman1996}}.
We adopt this empirically derived value for our ionospheric model.

To generate a turbulent screen in LoSiTo, the algorithm described in \cite{buscher2016} is employed. This algorithm follows an approach based on the fast Fourier transform method and treats large and small spatial frequencies separately to increase computational performance. 
Furthermore, it makes use of the frozen turbulence approximation. In this approximation, the change of the turbulent structure is assumed to be negligible with respect to the bulk velocity of the ionosphere. Therefore, the ionosphere is modeled as a static grid moving across the LOFAR stations. To optimize performance, the extent of the screen is constrained to the outermost pierce-points for each time-step. 
We scale the \dTEC sampled from the \TEC-screen such that the maximum \dTEC across all stations, directions and times is \SI{0.25}{\tecu}. We add a homogeneous component of the \TEC at \SI{7.0}{\tecu}. 
The daily ionospheric variation is included using a simple sinusoidal pattern: The total electron content peaks at 3\,p.m. and drops to a level of 10\% at 3\,a.m.  

We place our model-ionosphere at a height of $h_{ion}=\SI{250}{\kilo\meter}$, around the typical location of the peak electron density at the latitude of LOFAR as modeled by the international reference ionosphere \citep{IRI}. 
The ionospheric grid is simulated at an angular resolution of 1\,arcmin ($\approx\SI{72}{\meter}$) and moves from West to East at a velocity of \SI{20}{\meter\per\second}.

\begin{figure}
    \centering
    \includegraphics[width=0.99\linewidth]{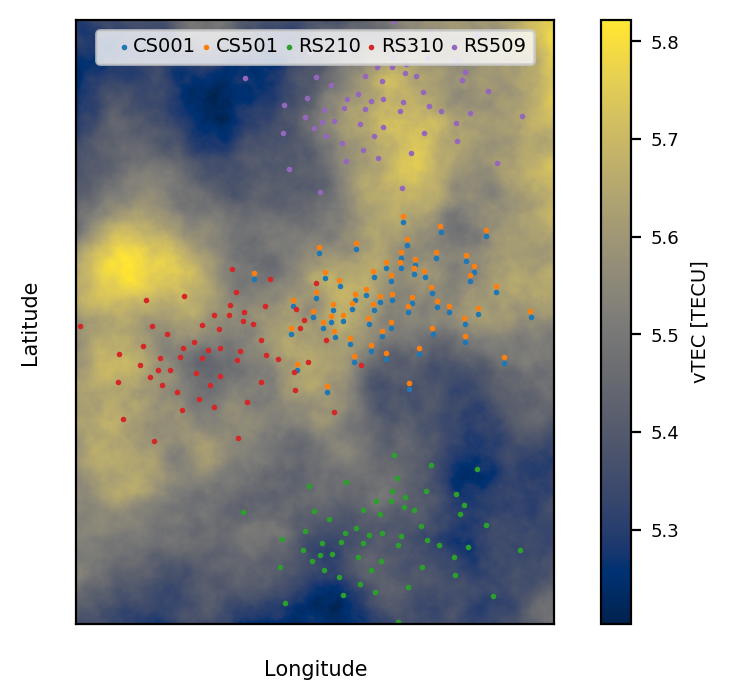}
    \caption{Snapshot of a simulated turbulent \TEC-screen with the parameters $\beta = 3.89$ and $h_{iono}=\SI{250}{\kilo\meter}$. Each circle marks one pierce-point, circles of the same color show pierce-points for different directions belonging to the same station. Only pierce-points belonging to one of the centermost stations (\emph{CS001}), the most distant central station (\emph{CS501}) as well as the three outermost remote stations (\emph{RS210, RS310, RS509}) are displayed.}
    \label{fig:tecscreen}
\end{figure}

One snapshot of such a \TEC-screen is shown in \autoref{fig:tecscreen}. We show the pierce-points of five selected LOFAR stations. It is apparent that all the CS observe a similar ionosphere, the \dTEC between different directions dominates compared to the \dTEC between different CS. The most distant RS observe a different part of the ionosphere. 

 For each pierce-point and at every time-step, the \vTEC-values of the closest grid points are linearly interpolated and converted to \sTEC according to \autoref{eq:sTEC}. These values are stored in the \texttt{h5parm} format introduced in \cite{deGasperin2019}.
 \begin{figure}
     \centering
     \includegraphics[width=0.99\linewidth]{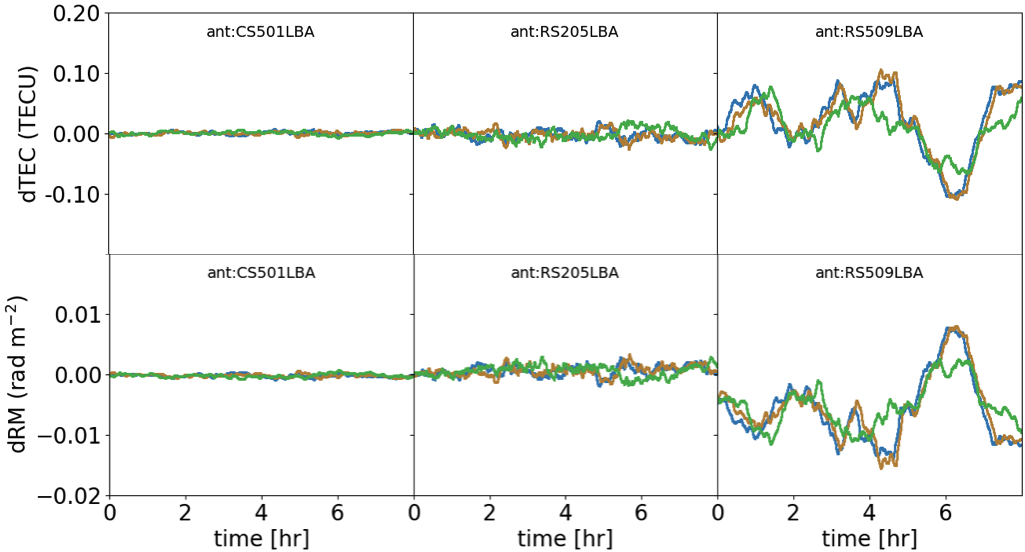}
     \caption{The top panels show simulated \dTEC for three stations towards the phase center (orange) and towards sources separated from the phase center by 1.75\textdegree\ (blue) and 4.51\textdegree\ (green). The stations are at distances of \SI{1.7}{\kilo\meter} (\textit{CS501}), \SI{6.3}{\kilo\meter} (\textit{RS205}) and \SI{55.7}{\kilo\meter} (\textit{RS509}) to \textit{CS001}. In the bottom panels, we show the corresponding simulated \dRM. The values of each direction are referenced with respect to the same direction in \textit{CS001LBA}.}
     \label{fig:simtecrm}
 \end{figure}
 
To derive the Faraday rotation from the ionospheric model, we apply the thin-layer approximation to \autoref{eq:rm}:
\begin{equation}
    \label{eq:frthinlayer}
    \RM = \SI{2.62e-13}{\per\tesla} \TEC \hat d\cdot\vec B\,\big\rvert_{\mathrm{pierce-point}}.
\end{equation} 
For calculation of the \RM, the magnetic field vector $\vec B$ projected onto the source-direction $\hat d$ and evaluated at the pierce-point is required. As magnetic field model, the implementation of the \emph{World Magnetic Model} \citep{chulliat2015} in the \texttt{RMExtract} library \citep{meviusRMextract} is employed. 

The \textit{FARADAY}-operation in \texttt{LoSiTo} derives the \RM from this magnetic model and the \sTEC values stored in the h5parm file. The \RM values are added to the h5parm-file in a new table.
The simulated \dTEC and \dRM-values for three stations and three directions are displayed in \autoref{fig:simtecrm}. As expected from \autoref{eq:frthinlayer},  the \dTEC and \dRM values are highly correlated. From our experience, the ionospheric conditions of this simulation are representative for low solar activity.

\subsection{The primary beam} 

The primary beam characterizes the directional sensitivity for the interferometer components, in the case of LOFAR for the individual stations. Since LOFAR does not simply consist of single dish antennas, but follows a phased array design, the treatment of the primary beam is more complex compared to classical interferometers.
For LOFAR, the primary beam can be split into two components, the \emph{element beam} and the \emph{array-factor}:
\begin{itemize}
    \item The \textit{element beam} characterizes the directional sensitivity of a single antenna element. An LBA element consist of one dipole pair, while for HBA, tiles of 16 pairs are grouped together to form one element \citep{vanHaarlem2013}. The element beam varies only slowly across the sky and in time. The element beam is always pointed towards the local horizon, this leads to a decline of sensitivity for low elevations. 
    \item The \textit{array-factor} is a Jones-scalar which describes the influence of the beam-forming, i.e. the superposition of the element signals, on the directional sensitivity. The array factor determines the size of the instrument FoV.
\end{itemize}
LoSiTo uses the \texttt{LOFARBeam}\footnote{\url{github.com/lofar-astron/LOFARBeam}} library to include both the element beam and the array factor in the simulations. This library uses results of electromagnetic simulations of the dipoles to calculate the primary beam Jones-matrices \citep{hamaker2011}.

\begin{figure}
    \centering
    \includegraphics[width=0.495\textwidth]{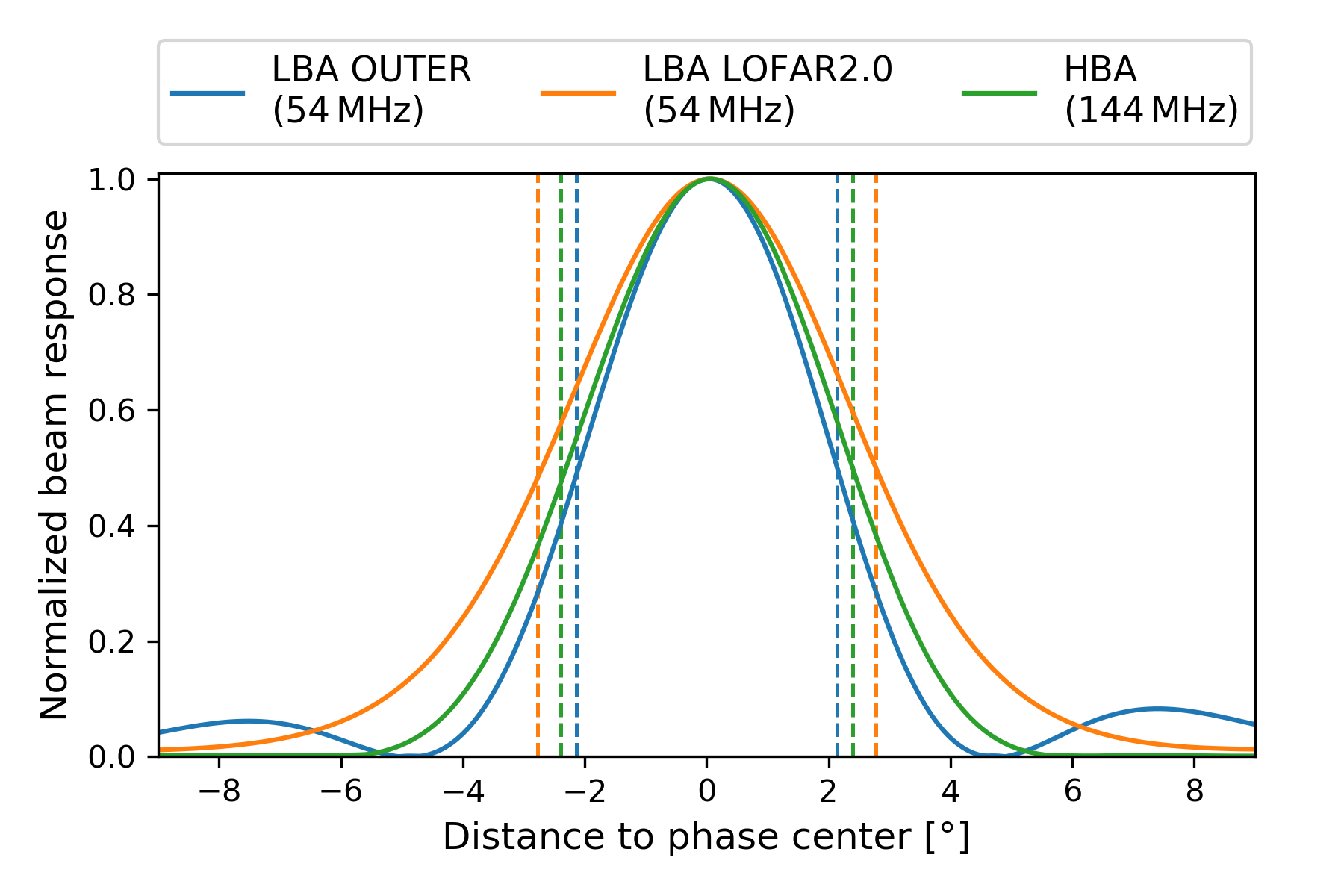}
    \caption{Normalized \texttt{LOFARBeam} beam response ($y$-axis) as a function of phase center separation ($x$-axis) for the current 48 dipole LBA system in mode \textit{outer}, the 96 dipole LOFAR\,2.0 LBA system and the HBA system. The dashed vertical lines mark the primary beam FWHM of the respective system.}
    \label{fig:beam}
\end{figure}

Since the LOFAR\,2.0 LBA system will utilize 96 dipoles per station instead of just 48, the shape of the array-factor will change. In \autoref{fig:beam}, we show a comparison of the array-factor amplitude between a current LBA station, a LOFAR\,2.0 LBA station and an HBA station. We note that the LOFAR\,2.0 upgrade will not just increase the size of the FoV, but also significantly reduce the side-lobe structure. The size of the primary beam at \SI{54}{\mega\hertz} (full-width at half-maximum) will increase to  5.6$^\circ$ from 4.3$^\circ$ in \textit{LBA\_outer}. For comparison, the HBA beam in mode \textit{dual\_inner} at \SI{144}{\mega\hertz} is 4.8$^\circ$ wide.

\subsection{Bandpass response}

The bandpass describes the frequency dependence of the system's amplitude response. Therefore, it is constant in time and does not affect phases. It is an instrumental effect which is largely shaped by the dipole response. In reality, variations in the dipole shapes and electronics cause deviations of the bandpass between the different stations and polarizations and might also introduce a slow time-variation.
This variation may be caused by changes in environmental conditions that affect dipole properties, such as humidity.
The bandpass can be described by a station dependent, real diagonal matrix.

\begin{figure}
\centering
    \includegraphics[width=1.0\linewidth]{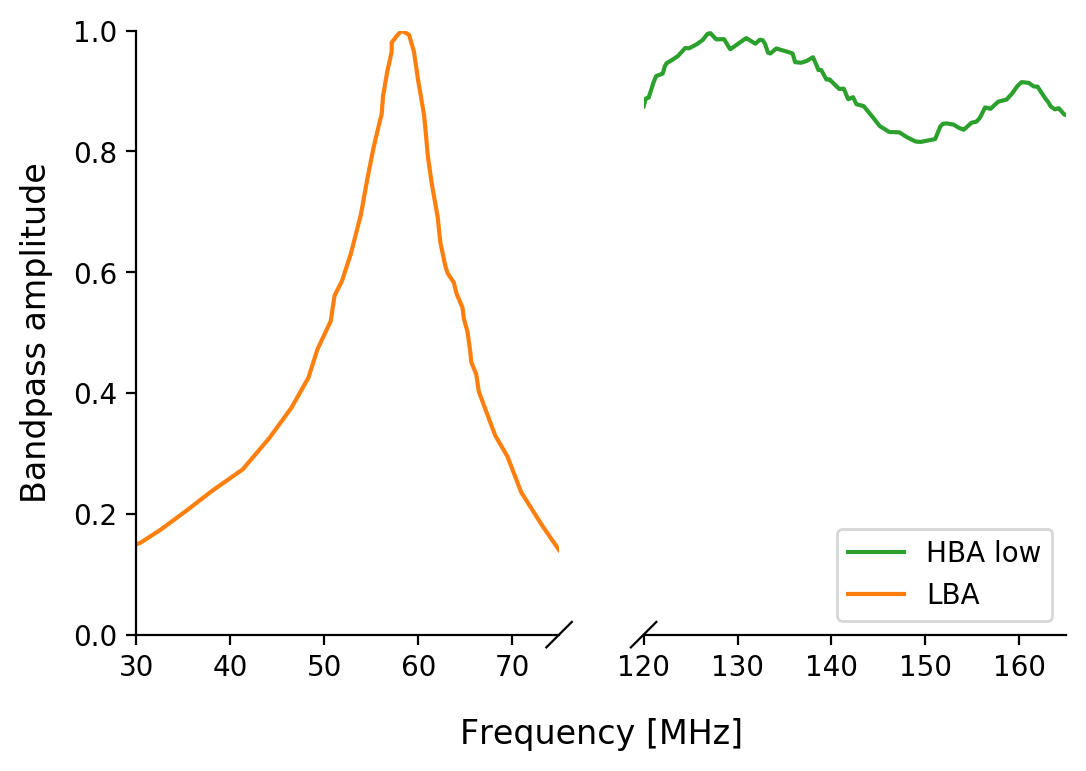}
    \caption{Normalized bandpass amplitudes ($y$-axis) for the LBA and the most common HBA bandpass setup \emph{HBA-low} as a function of frequency ($x$-axis). The values are adopted from an empirical study of the instrument in \citet{vanHaarlem2013}.} 
    \label{fig:bandpass}
\end{figure}

\texttt{LoSiTo} includes a straightforward model for the bandpass, using the average bandpass that was determined  in an empirical study of the instrument in \citet{vanHaarlem2013}. The model for both antenna types is shown in \autoref{fig:bandpass}. LOFAR features three HBA bandpass modes with different spectral windows. We will only consider the mode \emph{HBA-low}, since in practice, this mode is used almost exclusively. While the HBA dipole response is rather flat, the response of the LBA dipoles features a prominent peak at \SI{58}{\mega\hertz} with a full-width at half maximum (FWHM) of \SI{15}{\mega\hertz}, causing a sharp decline of sensitivity below and above the peak frequency.  

\subsection{Station based delays}

An omnipresent systematic effect in radio interferometry is given by offsets in the signal timing between the interferometer stations or antennas. This is for instance caused by asynchronous clocks and improper calibration of electronics. Especially for long baseline instruments, the issue of accurate clock calibration across large physical distances arises. 

A time delay $\Delta t$ between two baseline elements introduces a phase-offset of $\Delta\phi$ in the \textit{RIME}:
\begin{equation}
   \Delta\phi = 2\pi \nu \Delta t. 
\end{equation}
While this phase error follows a linear frequency dependence and is therefore a greater issue at higher frequencies, it is still non-negligible in present-day LOFAR observations. 
Currently, the remote and international stations receive a time signal from GPS-synchronized rubidium clocks \citep{vanHaarlem2013}. The signal of these clocks drifts slowly in time, causing a $\mathcal{O}(\SI{10}{\nano\second})$ clock error with smooth temporal behavior. This translates to a phase-error in the order of $2\pi$ at \SI{100}{\mega\hertz}. 
The LOFAR\,2.0 upgrade will introduce a new system for the time synchronization, utilizing a distributed signal from a single clock for all Dutch stations.
The requirements for the distributed single-clock signal are as follows (Bassa 2020, priv. comm.):
\begin{enumerate}
\item The timing reference signal at Core Stations shall have a
clock error of $\Delta t_{rms} < \SI{0.20}{\nano\second}$ over a 1-hour period.
\item The timing reference signal at Core and Remote Stations
shall have a clock error of $\Delta t_{rms} < \SI{0.35}{\nano\second}$ over an 8-hour
period.
\end{enumerate}
We use a simple model consisting of a sinusoidal variation on top of a constant offset to include the clock drift in the simulation:
\begin{equation}
    \Delta t_i = t_{\mathrm{amp},i}\sin\left(2\pi\omega_{\mathrm{clock},i} (t - t_{\mathrm{shift}})\right) + t_{\mathrm{offset}}.
    \label{eq:clock}
\end{equation}
The drift amplitude $t_{\mathrm{amp}}$, drift frequency $\omega_{\mathrm{clock}}$ and the clock offset $t_{\mathrm{offset}}$ are drawn from Gaussian distributions independently for each station. The widths $\sigma_\mathrm{offset}$ and $\sigma_\mathrm{amplitude}$ are determined such that the root-mean-square (\rms) error of the clock signal is one third of the maximum allowed \rms in the system requirements. Furthermore, the total standard deviation should be caused in equal parts by the constant offset and the sine function. The corresponding parameters are displayed in \autoref{tab:clockl2}.

\begin{figure}
    \centering
    \includegraphics[width=1.0\linewidth]{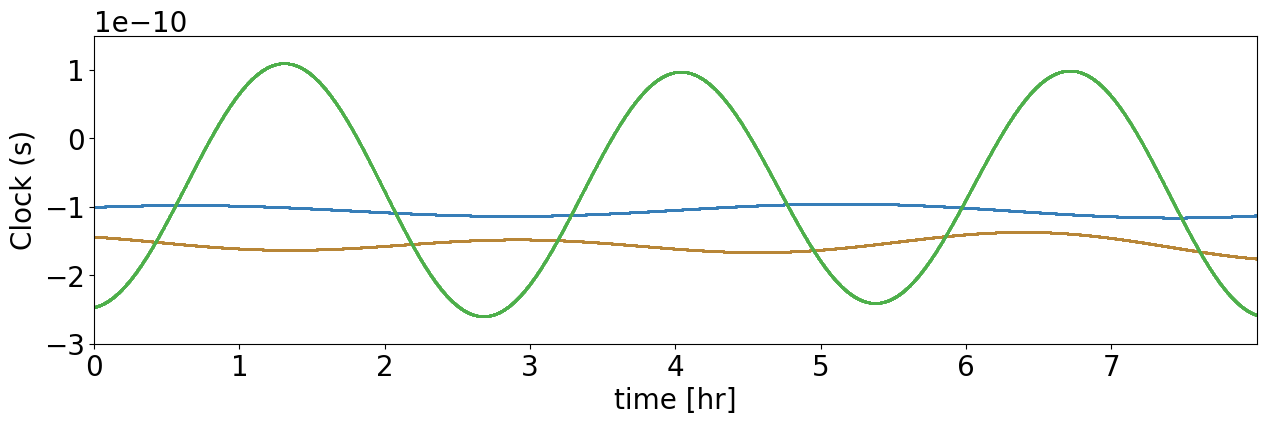}
    \caption{Simulated clock delay for \textit{CS501} (blue), \textit{RS205} (orange) and \textit{RS210} (green), referenced to the clock of \textit{CS001}. }
    \label{fig:simclock}
\end{figure}

\begin{table}
\caption[LOFAR\,2.0 clock model]{Numeric values for the widths of the Gaussian distributions used to construct the LOFAR\,2.0 clock model.}\label{tab:clockl2}
\begin{tabular}{ l c c c }
 Station type & $\Delta t_\mathrm{rms, max}$ [\si{\nano\second}] & $\sigma_\mathrm{offset}$ [\si{\nano\second}] & $\sigma_\mathrm{amplitude}$ [\si{\nano\second}] \\\hline
Core station & 0.20 & 0.047  & 0.067 \\  
Remote station & 0.35 & 0.083 & 0.117 \\
\end{tabular}
\end{table}
The resulting clock offset from this model is shown in \autoref{fig:simclock} for three selected stations. The corruptions are shared between the LBA and HBA parts of each station.

In addition to the station dependent clock delay, LOFAR data shows the presence of a nanosecond-scale time delay between the $X$ and $Y$ polarization of the station output. This delay, referred to as polarization misalignment, is constant in time and attributed to an inaccurate station-calibration \citep{deGasperin2019}. We assume that this effect will be present at a similar magnitude in LOFAR\,2.0. 
To replicate this  effect, a random time offset between the polarizations is drawn from a Gaussian distribution with a width of \SI{1}{\nano\second} for each station. We sampled this effect independently for the LBA and HBA parts of a station.

\subsection{Thermal noise}

The achievable sensitivity of a perfectly calibrated radio interferometer is limited by the presence of noise. There are two primary noise sources: instrumental noise, for example from the receiver and amplifier system, and sky noise, which is of cosmic origin. While for radio observations at mid and high frequencies, the instrumental component is by far dominant, this changes in the low-frequency regime, were the sky noise becomes increasingly important. 

The sky noise has an ultra-steep spectrum. Assuming a power-law shape $I_{sky} \propto \nu^{-\alpha}$, the study in \cite{guzman2011} found a spectral index of $\alpha = 2.57$ between \SI{45}{\mega\hertz} and \SI{408}{\mega\hertz}.  
This spectral index varies across the sky, increasing strongly towards the Galactic plane. Towards the Galactic center, the brightness of the sky noise is a factor of $\approx$ 10 higher compared to the Galactic poles. 
For LOFAR, the sky noise is in fact the dominant source of noise below \SI{65}{\mega\hertz} \citep{vanHaarlem2013}. 

\begin{figure*}
    \centering
    \includegraphics[width=0.49\linewidth]{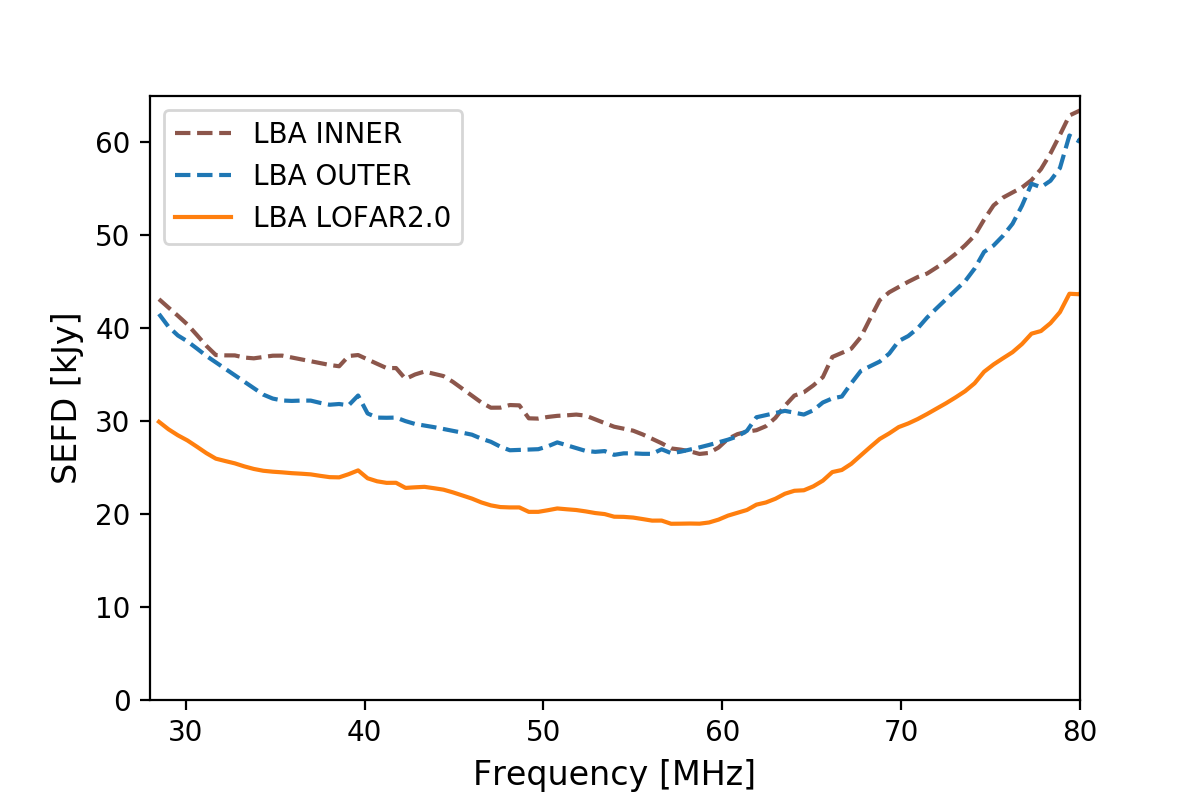}
    \includegraphics[width=0.49\linewidth]{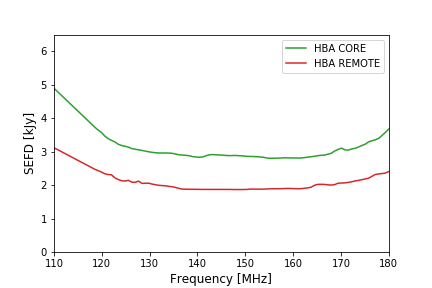}
    \caption[Source equivalent flux density]{Frequency dependence of the \SEFD  for different configurations of the LBA (left) and two differently equipped HBA station types (right). The estimate for the LOFAR\,2.0 LBA \SEFD is derived from the mean of the modes \textit{Inner} and \textit{Outer} and scaled by a factor of 0.71 to account for the increased dipole count.}
    \label{fig:noise}
\end{figure*}

The noise of a radio astronomical instrument can be expressed in terms of system equivalent flux density (\SEFD). The \SEFD is the flux density of a hypothetical source which induces a power in the system that is equal to the power induced by the noise. It can be calculated from the system temperature $T_\mathrm{sys}$, the antenna efficiency $\eta$ and the effective collection area $A_\mathrm{eff}$:
\begin{equation}
    \SEFD = \frac{2k_\mathrm{B}T_\mathrm{sys}}{\eta A_\mathrm{eff} }.
    \label{eq:sefddef}
\end{equation}
An empirical study in \citet{vanHaarlem2013} determined the \SEFD  for the LOFAR LBA and HBA. For the LBA system, the \SEFD was measured independently for the two observation modes \textit{outer} and \textit{inner}, where only the 48 inner- or outermost of the 96 dipoles are used. To estimate the \SEFD for the LOFAR\,2.0 LBA, we take the mean of the two modes as reference. Additionally, we scale this mean value to account for the double dipole number. The anti-proportionality to the effective collection area in \autoref{eq:sefddef} suggests that a doubling of the dipoles reduces the \SEFD by a factor of 2. However, we adopt a more conservative scaling factor of $\frac{1}{\sqrt2}\approx0.71$ for two reasons:
First, overlap in the effective areas of the dipoles leads to a decrease in the total area. Second, we compare the \SEFD of the different HBA station types displayed in \autoref{fig:noise}. The HBA arrays in the CS host 24 tiles, while the RS host 48 tiles, so they can give an idea on how a doubling of the dipoles affects the \SEFD. The average \SEFD of the 48 tile stations is lower by a factor of 0.75. Due to the different station layouts, this ratio cannot be transferred to LBA directly. Nevertheless, this comparison motivates our slightly more conservative value of 0.71. Our estimate for the LOFAR\,2.0 LBA \SEFD as well as the values adopted from \citet{vanHaarlem2013} are displayed in \autoref{fig:noise}.

The noise in visibility space follows a Gaussian distribution with a frequency dependent standard deviation. This standard deviation for a frequency channel of bandwidth $\Delta\nu$ can be computed from the \SEFD of the two stations which form the baseline, the total system efficiency $\eta_{sys}$ and the exposure $\tau$ \citep{taylor1999}:
\begin{equation}
    \Delta S_{ij}(\nu) = \frac{1}{\eta_{sys}} \sqrt{\frac{\mathrm{SEFD_i(\nu)SEFD_j(\nu)}}{2\Delta\nu\tau}}.
\end{equation}
We assume a system efficiency of $\eta_{sys}=0.95$. 
Using these standard deviations, the simulated complex visibilities are corrupted with independent Gaussian noise in real and imaginary part for each frequency channel, baseline, time and polarization.

\section{Simulation software}

\label{sec:losito}
\begin{figure}
    \centering
    \includegraphics[width=\linewidth]{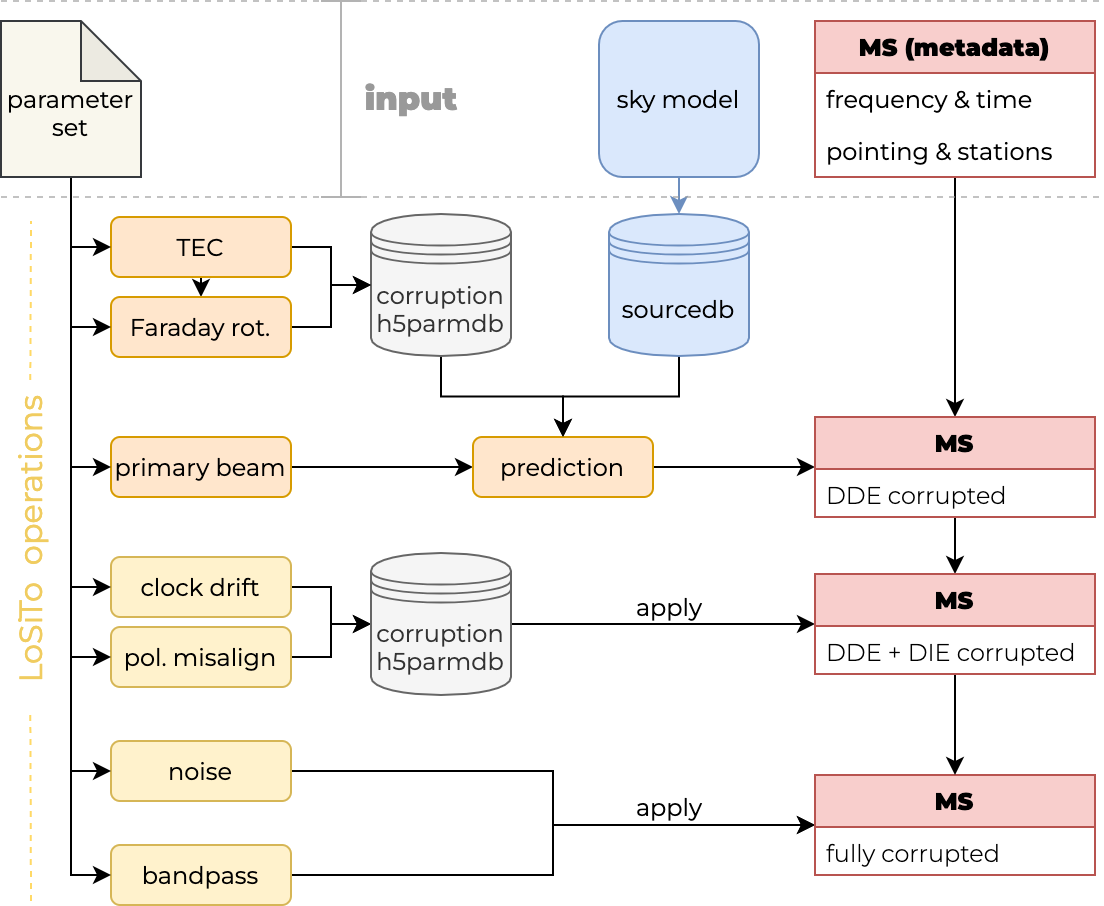}
    \caption{Diagram of \texttt{LoSiTo}. As input, a parameter set, a sky model and a template measurement set are required. Orange boxes represent DDE, while yellow boxes represent DIE.}
    \label{fig:losito}
\end{figure}

We implemented the models for the corrupting effects presented in \autoref{sec:corruptions} in a code called the \emph{LOFAR simulation tool} (\texttt{LoSiTo})\footnote{\url{github.com/darafferty/losito}}. This software is a command-line program written in the Python programming language and build on top of existing LOFAR software. 
In the following, we provide a brief overview of the program.
The main configuration file is the \emph{parameter set} in which the user specifies which corrupting effects should be include in the simulation and at which scale. Two more input files are required: one is the input sky model, where properties of the sources such as position, flux density, angular extension, spectral shape and polarization properties are set. Such a sky model may be obtained from a source catalog or a radio image.  Alternatively, it can be randomly generated using a script in \texttt{LoSiTo}. As last input for the simulation, a measurement set file \citep{MS_lofar} is required. This file is the template for the simulated visibilities, furthermore it stores the metadata of the observation, such as the observation time, frequency bands as well as location and status of the LOFAR stations. 
A \texttt{LoSiTo} simulation is composed of individual operations, each model corruption of \autoref{sec:corruptions} is implemented as one such operation. The simulated corruptions are stored in the \texttt{h5parm} data format. The central part of a simulation is the prediction. In this step, the visibilities corresponding to each source (or each patch of sources) are calculated in a Fourier-transformation from image-space to visibility-space. The Jones-matrices of the direction-dependent effects (DDE), such as the ionospheric effects and the primary beam, are calculated from the \texttt{h5parm} file content and multiplied with the predicted visibility matrices for the source(s). The resulting DDE corrupted visibilites for all directions are added.
The Jones-matrices of the direction-independent effects (DIE) are multiplied with the visibility matrices afterwards to save computation time. This is possible because the DDE are prior to the DIE on the signal path. Lastly, noise is added to the visibilities and they are multiplied by the average bandpass response. The final product of the simulation is a measurement set file which obeys the same format as a real LOFAR observation, and thus, can be further processed with the same software. \texttt{LoSiTo} makes use of the software \texttt{DPPP} for the prediction and the application of the corruptions stored in \textit{h5parm} files \citep{software_DP3}.
The diagram in \autoref{fig:losito} visualizes the architecture of \texttt{LoSiTo}. 

\subsection{Full simulation setup}

We simulate a full 8-hour LOFAR\,2.0 observation of a calibrator source and a target field using the Dutch LBA and HBA stations simultaneously. For the LBA system, multi-beam observing allows to point in parallel at both target and the calibrator during the whole observation. This is not possible for HBA, were we simulate a short calibrator scan of 10 minutes at the beginning of the observation. The setup of the simulated observation is shown in \autoref{tab:metadata}. Since target and calibrator field are usually located at large angular separation, they were simulated using independent ionospheric models. The station-dependent instrumental effects of bandpass, clock and polarization alignment are shared between the calibrator and target data set. 
\begin{table}
\centering
\caption{Setup of the simulated observation. }\label{tab:metadata}
\begin{tabular}{ l c }\hline
Start time & 2018-06-30 11:00:02  \\
Observation time & \SI{8}{\hour}\tablefootmark{a}  \\  
Time resolution & \SI{4}{\second} \\
Frequency range (LBA) & 30-\SI{78}{\mega\hertz} \\
Frequency range (HBA) & 120-\SI{168}{\mega\hertz} \\
Frequency resolution & \SI{48.8}{\kilo\hertz} \\
Target RA & 10$^\mathrm{h}$:31$^\mathrm{m}$:41$^\mathrm{s}$ \\
Target Dec & +35$^\circ$04'38'' \\
Calibrator RA & 08$^\mathrm{h}$13$^\mathrm{m}$35$^\mathrm{s}$ \\
Calibrator Dec & +48$^\circ$13'02''\\\hline
\end{tabular}
\tablefoot{
\tablefoottext{a}{Ten minutes for the HBA calibrator scan.}}
\end{table}

\begin{figure}
    \centering
    \includegraphics[width=\linewidth]{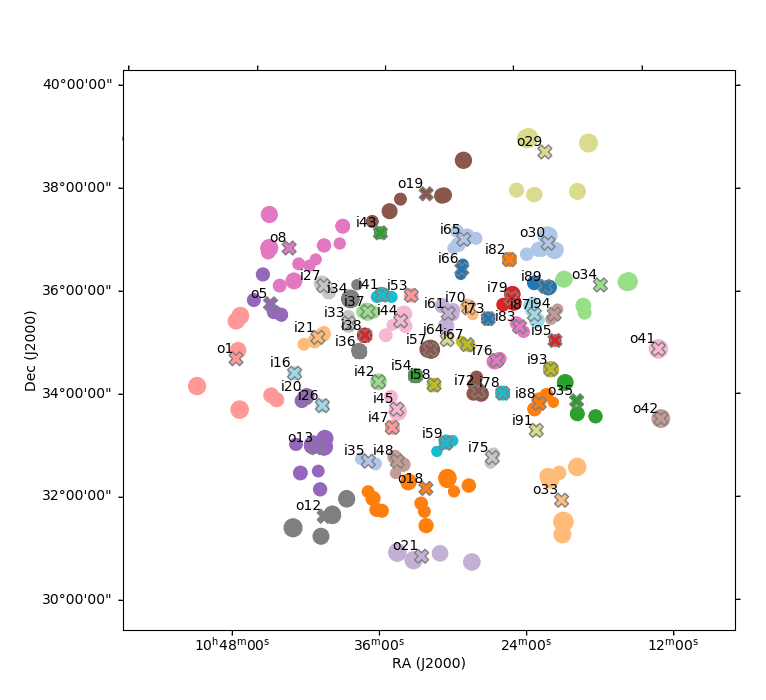}
    \caption{Input sky model of the simulation. The model is extracted from a real LOFAR observation of the target field. It contains 340 sources grouped into 59 patches, the total flux density is \SI{226}{\jansky} at \SI{56}{\mega\hertz}. Circles correspond to source locations. Sources are color-coded according to their patch membership, patch centers are marked by an \emph{`X'} and labeled.}    \label{fig:skymodel}
\end{figure}

We extracted the sky model for the target from a real LOFAR LBA observation of the field around the galaxy cluster \emph{A1033} using the software \texttt{PyBDSF} \citep{mohan2015}. We corrected the sky model for the primary beam attenuation and set the spectral shape of all sources to a power law with a typical radio-spectral index of 0.8 \citep{mahony2016}. We manually adjusted the flux density of the bright, resolved source in the center of the field to make it less dominant. Furthermore, we discarded all sources fainter than \SI{0.2}{\jansky} to decrease computational cost. Lastly, to further improve computational efficiency, we grouped sources in close proximity into patches. The DDE are not applied to each individual source but to each patch of sources, taking the flux-weighted centroid of the source locations as reference. Further away than 2.5\textdegree\ from the phase center, sources are grouped into larger patches. The resulting sky model is displayed in \autoref{fig:skymodel}, it contains 340 sources in 59 patches, the total flux density is \SI{225.8}{\jansky} at \SI{54}{\mega\hertz}. 
For the calibrator observation, we use an existing model of the standard LOFAR calibrator source \textit{3C196} which has a total flux density of \SI{138.8}{\jansky} at \SI{54}{\mega\hertz}.
We create the input measurement sets using a script available in \texttt{LoSiTo}. This way, we receive measurement sets for LBA and HBA with identical time-steps and pointing information. To take into account the changes introduces in the LOFAR\,2.0 upgrade, we set all 96 LBA dipoles for each station to active. 

We ran the simulation including all the corruptions discussed in \autoref{sec:corruptions}. The full computation time for the simulation was 53 hours using eight compute nodes equipped with \textit{Intel Xeon E5-1650} six-core processors. The primary bottleneck in terms of computation time is the prediction which depends on the number of patches and sources present in the sky model. The simulated data is made publicly available\footnote{\url{https://www.fdr.uni-hamburg.de/record/8587}}.

\section{Calibration}

\label{sec:calibration}
Simultaneous LBA and HBA observations will offer new prospects for ionospheric calibration.  Since the same region of the ionosphere will be observed in both parts of the array, the underlying parameters describing the ionospheric corruption in the data are identical. Consequently, combining the information of the low- and high-band observation could allow one to determine the ionospheric parameters more accurately and hence, derive more exact calibration solutions. This strategy is possible since both dipole systems have a comparable primary beam size which allows one to observe a sufficient number of calibrator sources in both parts of the array. It could especially benefit the LBA where calibration of DDE is harder due to the increased noise level and severity of ionospheric errors.

Currently, calibration of DDE for LOFAR mostly follows a brute-force attempt: it is solved for effective Jones matrices which are assumed to be constant for small domains of time and frequency \citep{Tasse2021,deGasperin2020}. However, to exploit the simultaneous observation, parameters describing the underlying physical effects, such as \TEC and \RM, must be obtained, since only these values are frequency-independent and can be meaningfully translated between LBA and HBA. This drastically reduces the number of free parameters during calibration compared to the effective-Jones approach, but requires a sufficient signal-to-noise ratio and data which is clean of other systematic errors to converge towards the correct solutions. 

We focus on two approaches to exploit the simultaneous observation for calibration of the direction-dependent \dTEC. One approach is the application of ionospheric solutions found in HBA to the LBA data, we will refer to this method as \emph{solution transfer}. This idea is based on the assumption that it is easier to converge towards correct calibration solutions in HBA, since there is significantly less noise compared to LBA (see \autoref{fig:noise}). However, while the HBA stations are less affected by noise, they also observe a smaller fractional bandwidth ($\approx 0.33$) compared to LBA ($\approx0.89$) which could make it harder to fit the $\nu^{-1}$ spectral dependency (see \autoref{eq:dispersive_delay}) of the dispersive delay. Therefore, it is not fully clear how much more accurate \TEC-extraction is in HBA. In addition, we emphasize that, when applying HBA-derived solutions to LBA, it is crucial to minimize the presence of jumps in \dTEC solutions. These jumps are a consequence of the local minima present in the $\chi^2$ cost function which may be almost as deep as the global minimum, depending on the bandwidth \citep{vanWeeren2016}. While for HBA, jumps to a local minimum might still provide accurate calibration for a significant part of the HBA bandwidth, they cause substantially stronger phase-errors when transferred to LBA since the location of the local minima shifts with frequency.

The second method of exploiting the simultaneous observation we consider in our analysis is  \emph{joint calibration}. In this scenario, it is solved for \dTEC  using data from both LBA and HBA together to leverage the large bandwidth observed by LOFAR and improve the signal-to-noise ratio. This approach contains more information on a larger domain and thus, carries a greater potential for calibration, although at the cost of a more complex algorithm.

The data reduction of the simulated observation is split into three parts, if not stated otherwise, data reduction steps are carried out independently for the LBA and HBA observations. First, the instrumental systematics are derived from the calibrator observation using the strategy described in \citet{deGasperin2019}. The second step is the calibration of  DIE employing an adjusted version of the pipeline presented in \cite{deGasperin2020}. The final step is the direction-dependent calibration, here we compare the LOFAR\,2.0-specific calibration strategies to the independent calibration of the LBA system.
 
\subsection{Calibrator data reduction}
\label{sec:cal-calib}

\begin{figure*}
\centering
\includegraphics[width=0.8\linewidth]{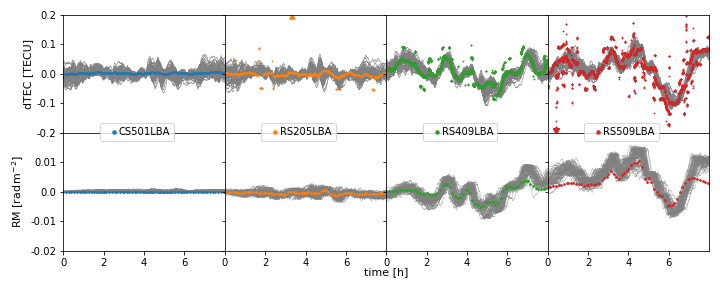}
\caption{Direction-independent ionospheric solutions towards the target field for four LBA stations. The top panels show \dTEC, the bottom panels \dRM. The gray curves in the background show the input corruptions towards the 59 patches. The calibration solutions are referenced to station \textit{CS001LBA} and the input corruptions are referenced to the phase center direction of \textit{CS001LBA}. Triangles mark solutions outside of the graph's scale. Note: The missing variation of the input-\RM for \textit{CS001} is caused by a referencing error present in the data. Simulated \RM values were referenced for each direction individually instead of referencing to a single direction. However, we do not expect this to affect our analysis.}\label{fig:soldielba}
\end{figure*}
\begin{figure*}
\centering
\includegraphics[width=0.8\linewidth]{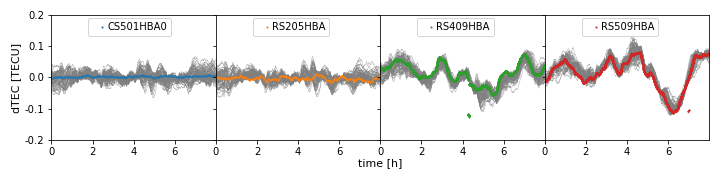}
\caption{Direction-independent \TEC solutions for four HBA stations. The gray curves in the background show the input corruptions towards the 59 patches. The calibration solutions are referenced to station \textit{CS001HBA0}. The input corruptions are referenced to the phase center direction of \textit{CS001HBA0}.}\label{fig:soldiehba}
\end{figure*}
\begin{figure}
    \centering
    \includegraphics[width=0.99\linewidth]{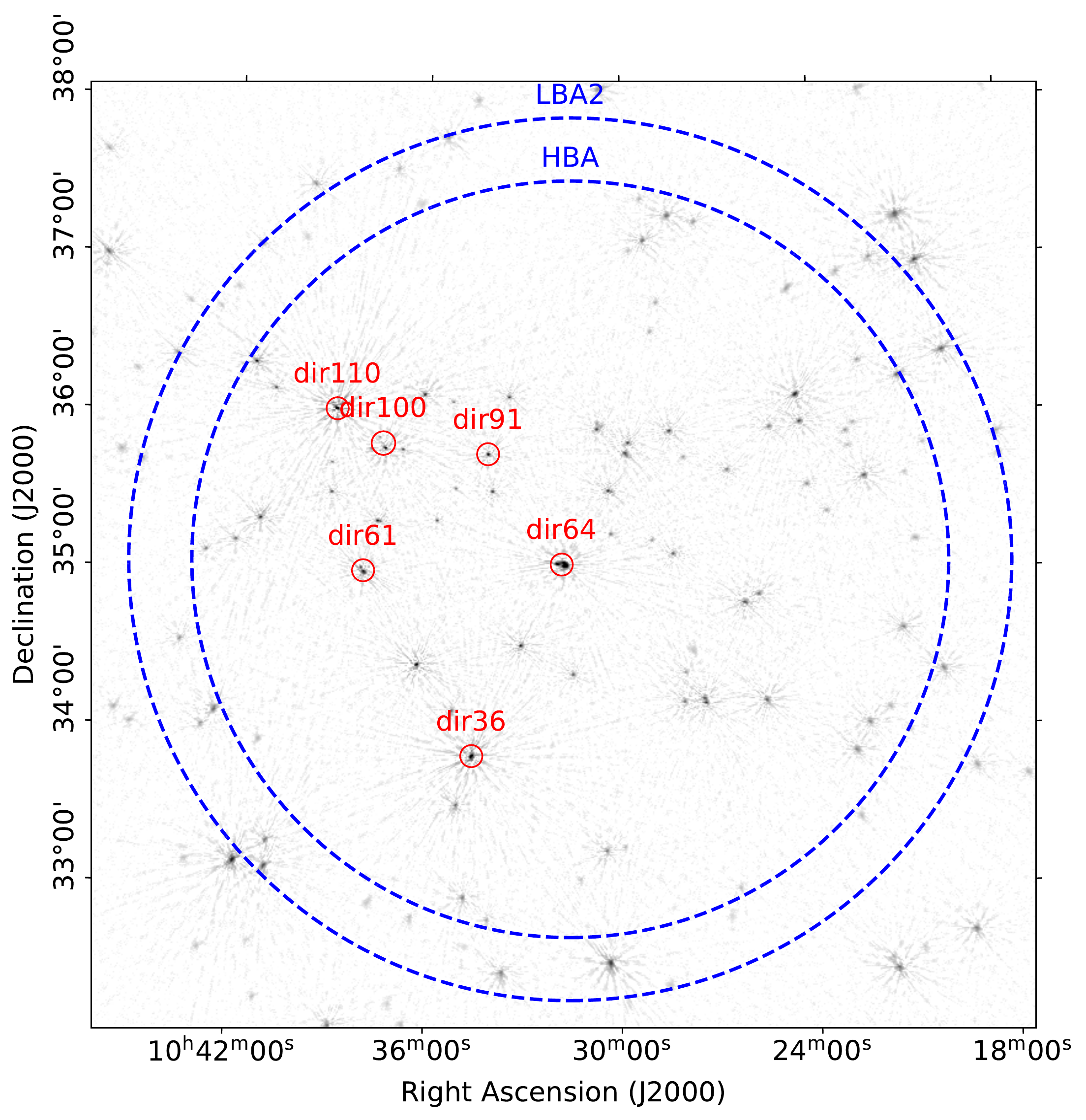}
    \caption{LBA wide-field image after DIE calibration. The \rms background noise is $\approx\SI{2.0}{\milli\jansky\per\beam}$ at a resolution of $42'' \times 30''$. The DDE calibrator directions are highlighted in red, the blue dashed circles show the primary beam $\mathit{FWHM}$ of the LOFAR\,2.0 LBA at \SI{54}{\mega\hertz} and the HBA at \SI{144}{\mega\hertz}. The background color scale shows the logarithmic surface brightness in arbitrary units.}
    \label{fig:caldirs}
\end{figure}

To find the solutions for the calibrator observation, we assume that the model of the calibrator source is fully accurate and calibrate against the same model we used as simulation input. 
First, solving against this model, the polarization alignment was derived by fitting a diagonal and a rotational matrix. The phase solutions from the diagonal matrix were averaged in time to extract the relative delay between the polarizations for each station.  We corrected the data for these delays. Next, the LOFAR beam model was applied and a second diagonal and rotational fit was performed to extract the Faraday rotation solutions. The Faraday rotation solutions are applied to the data. Subsequently, a diagonal Jones-matrix was fitted and the resulting amplitude solutions were averaged in time to derive the bandpass responses.
From here, the calibrator pipeline in \cite{deGasperin2019} performs additional solution steps to derive the time-dependent clock delay from phase solutions. In a procedure called  \emph{clock-TEC-separation}, a clock term proportional to $\nu$ and a \TEC term proportional to $\nu^{-1}$ are fitted to the phase solutions simultaneously to separate the effects by their characteristic frequency dependence \citep{vanWeeren2016}. For HBA, calibrator observations are usually not simultaneous to the target observation due to limitations imposed by the tile beam. Therefore, we skip this step for the HBA data set. For LBA, additional smoothing was required to account for the significantly smaller clock error of our LOFAR\,2.0 clock model.
A benefit of testing the data reduction strategy on simulated data is that we can quantify the accuracy of the calibration solutions: for LBA, the frequency-averaged relative \rms-error of the bandpass solutions is 1.8\% and the \rms error of the polarization alignment and clock delay is \SI{7.7}{\pico\second} respectively \SI{37}{\pico\second}. 
For HBA, the frequency-averaged relative \rms error of the bandpass solutions derived from the 10 minute calibrator observation is 0.3\% and the \rms error of the polarization alignment delays is  \SI{1.9}{\pico\second}.
The bandpass, polarization alignment and clock solutions were transferred to the target data and the primary beam in direction of the phase center was applied. We exploit the simultaneity of the LOFAR\,2.0 observations and transfer the LBA clock solutions to the HBA observation.

\subsection{Direction-independent calibration}
\label{sec:die}
Next, we calibrated the target field data sets, starting with direction-independent self-calibration based on the procedure described in \citet{deGasperin2020}. The aim of this is to correct for the average ionospheric effects per station and to derive a robust source model for further direction-dependent calibration. For real observations, a catalog model of the target field is used as initial model for self-calibration. To replicate this incomplete first model, we use a sparse, corrupted version of the simulation input sky model. This model was obtained by selecting only sources within the primary beam $\mathit{FWHM}$ which are brighter than 0.5\,Jy at \SI{54}{\mega\hertz}. Flux density errors at a standard deviation of 10\% were introduced to the remaining sources, which were 70 for LBA respectively 39 for HBA. First, we find solutions for the differential Faraday rotation: we transform the data to circular polarization basis and consider only the phase differences of the $XX$ and $YY$ correlations. This eliminates all contributions of scalar phase errors such as \TEC or clock drift since for unpolarized sources, they are equally present in both diagonal entries and cancel out. We fit these circular phase differences to a model with a phase of $\phi = 0$, extract the \RM from the phase solutions and apply it to the data. Next, the pipeline solves for the direction-averaged \dTEC in two sub-steps: first, only for the CS and the inner RS, these solutions are then applied to the data. Second, only for the outer RS, constraining all other stations to the same value. This improves the S/N ratio when determining the large \TEC-variations of the most distant stations. The solver estimates the \dTEC by fitting the $1/\nu$ term of the dispersive delay phase error.
In \autoref{fig:soldielba}, the direction-independent \dTEC and \RM solutions for four LBA stations are displayed on top of the corresponding input corruptions. The direction-independent solutions trace a weighted average of the input corruptions towards the different directions. For the most distant stations, the solver sometimes converged towards neighboring minima due to the presence of noise. The \RM solutions are less noisy since we chose a significantly longer solution interval of \SI{8}{\minute} instead of \SI{4}{\second}, however, for a few stations (such as RS509), they sometimes show a systematic deviation from the input corruptions.

The HBA \dTEC solutions are shown  in \autoref{fig:soldiehba}; due to the lower noise level, they have significantly fewer jumps. 
The corrected data is imaged using the \texttt{WSCLEAN} multi-scale algorithm \citep{Offringa2014, Offringa2016}. The \textit{CLEAN}-components found during imaging are used as improved model for direction-dependent calibration.

 For LBA, the image is displayed in \autoref{fig:caldirs}, the \textit{rms} background noise is $\SI{2.0}{\milli\jansky\per\beam}$ at a resolution of $42'' \times 30''$, for HBA, the noise level is  $\SI{200}{\micro\jansky\per\beam}$ at a resolution of $18'' \times 14''$. For comparison, typical values for the \rms background noise of real LOFAR observations are around $5 \si{\milli\jansky\per\beam}$ for LBA \citep{deGasperin2021} respectively $\SI{380}{\micro\jansky\per\beam}$ for HBA \citep{shimwell2019}. The lower noise can be attributed to the improved sensitivity of LOFAR\,2.0 for LBA and a reduced source density of our simulated sky model compared to a fully realistic source distribution. 

One point that must be accounted for in LOFAR\,2.0 calibration is the difference in FoV of the low- and high band (see \autoref{fig:caldirs}). The LBA beam at the center of the frequency band covers a $36\%$ larger area compared to HBA. This discrepancy could be compensated for by using multiple simultaneous HBA pointings to cover one LBA pointing. Alternatively, the greater sensitivity of the HBA could justify to use regions outside of the \textit{FWHM} for simultaneous calibration. 

\subsection{Direction-dependent calibration}

The last data-reduction step is the calibration for DDE which is necessary to correct for \TEC-variation across the FoV. We identify DDE-calibrators from the DIE calibrated LBA image, using the source finder \texttt{PyBDSF} to isolate islands of emission. We employ a grouping algorithm to merge sources in close proximity. To avoid faint and very extended sources, we place a threshold on the flux density to source area ratio and discard all sources fainter than \SI{0.8}{\jansky} at \SI{60}{\mega\hertz}. This procedure resulted in six calibrator directions with apparent flux densities of 0.9 to 2.8\,Jy, the locations of which are shown in \autoref{fig:caldirs}. We note that for sufficient calibration of the full FoV, more calibrator directions are necessary, depending on the ionospheric conditions. However, our sample contains point-like, multi-component and complex sources, and can thus give a good indication of the convergence for different morphologies. Expanding our strategy to more directions is straightforward as long as they are sufficiently bright.

\begin{figure}
\centering
\begin{subfigure}{\linewidth}
   \includegraphics[width=1\linewidth]{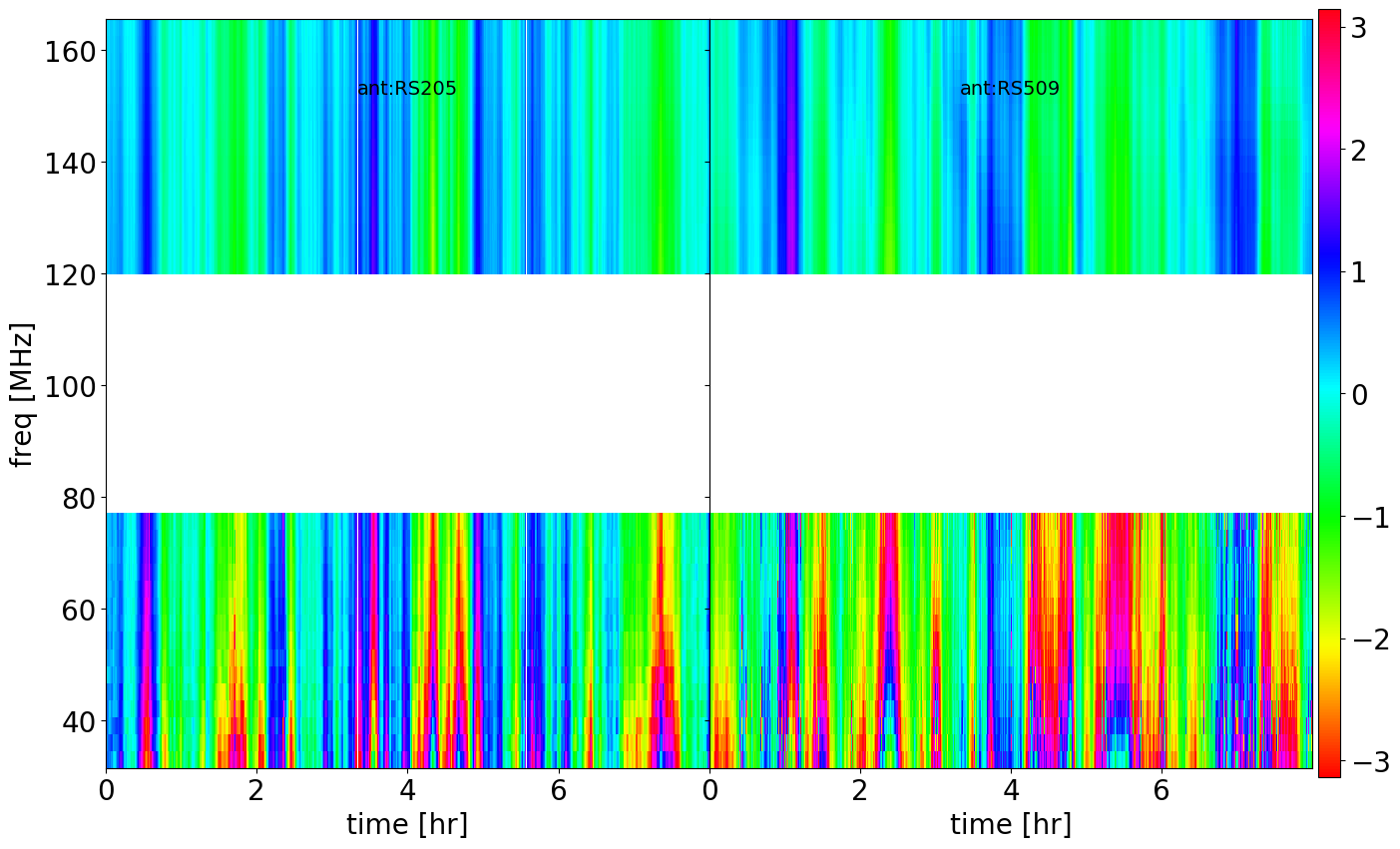}
   \caption{Phase solutions towards \textit{dir36}.}
\end{subfigure}
\begin{subfigure}{\linewidth}
   \includegraphics[width=1\linewidth]{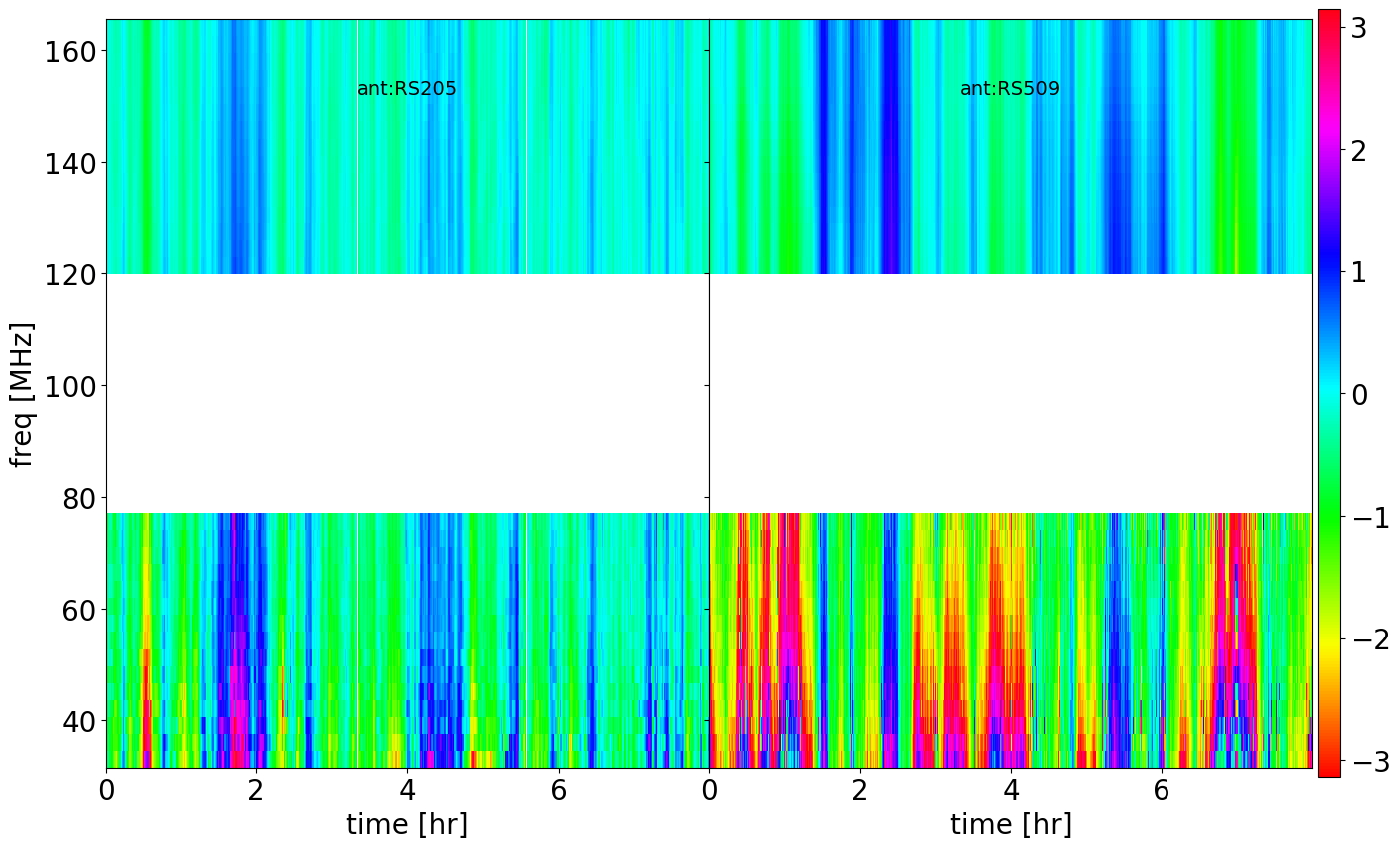}
   \caption{Phase solutions towards \textit{dir100}.}
\end{subfigure}
\caption{Phase solutions $\phi$ in radians in LBA and HBA for the two stations \textit{RS205} and \textit{RS509} as a function of time ($x-$axis) and frequency ($y$-axis). The top figure (a) shows solutions towards calibrator direction \emph{dir36}, the lower figure (b) towards \emph{dir100}.}\label{fig:ddeph}
\end{figure}
We pursue an approach based on the peeling-strategy \citep{vanWeeren2016}: We start by time-averaging the DIE-corrected data set to a resolution of \SI{8}{\second} and subtract all \texttt{CLEAN}-components from the DIE calibrated image. This creates a data set which is empty up to model inaccuracies and calibration residuals. We then iterate on our calibrator sources, starting with the brightest direction. We add the visibilities corresponding to the model of sources in this calibrator direction back to the measurement set. We create a measurement set for this specific direction by phase-shifting the data to the calibration direction and further averaging a factor of four in time and eight in frequency. On this data, we estimate the direction-dependent Faraday rotation from the circular-base $XX-YY$ phase difference as described in \autoref{sec:die}. After applying these solutions, we perform several rounds of self-calibration, solving for scalar phases in time intervals of down to \SI{32}{s} for each channel. We employ a \emph{station constraint}, forcing all core stations to the same solution, for these stations, direction-dependent variation of the ionosphere is negligible. The resulting phases are smoothed, using a Gaussian kernel with a standard deviation of \SI{5}{\mega\hertz} at \SI{54}{\mega\hertz}. The kernel-size varies as $\nu^{-1}$ in frequency to allow for more smoothing in frequencies less affected by the ionospheric errors. The self-iteration loop is discontinued once the \rms background noise of the calibrator region reduces for less than one percent.
The phase solutions and source model of the iteration with the lowest background \rms are used to re-subtract the calibrator sources from the data, reducing remaining artifacts. This improved empty data set is used as starting point for the next calibrator direction.
This cycle is repeated for all eight calibration directions for both LBA and HBA. The resulting phase solutions for two stations and two directions are shown in \autoref{fig:ddeph}. Temporal correlation of the solutions between LBA and HBA is visible, they are dominated by ionospheric DDE. 

\begin{figure*}
\centering
\begin{subfigure}{0.495\linewidth}
   \includegraphics[width=1.0\linewidth]{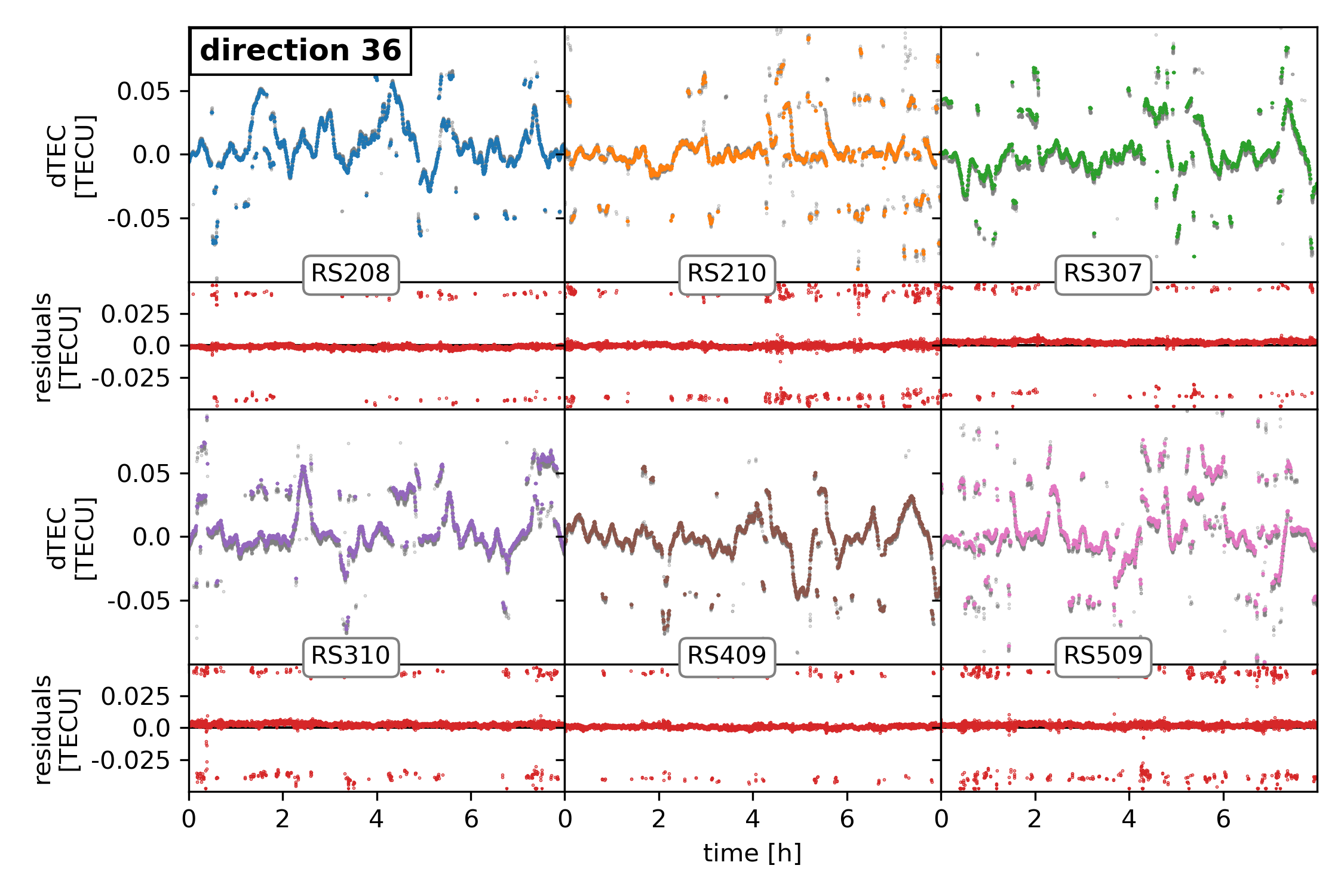}
   \caption{LBA - direction 36.}\label{a}
\end{subfigure}
\begin{subfigure}{0.495\linewidth}
   \includegraphics[width=1.0\linewidth]{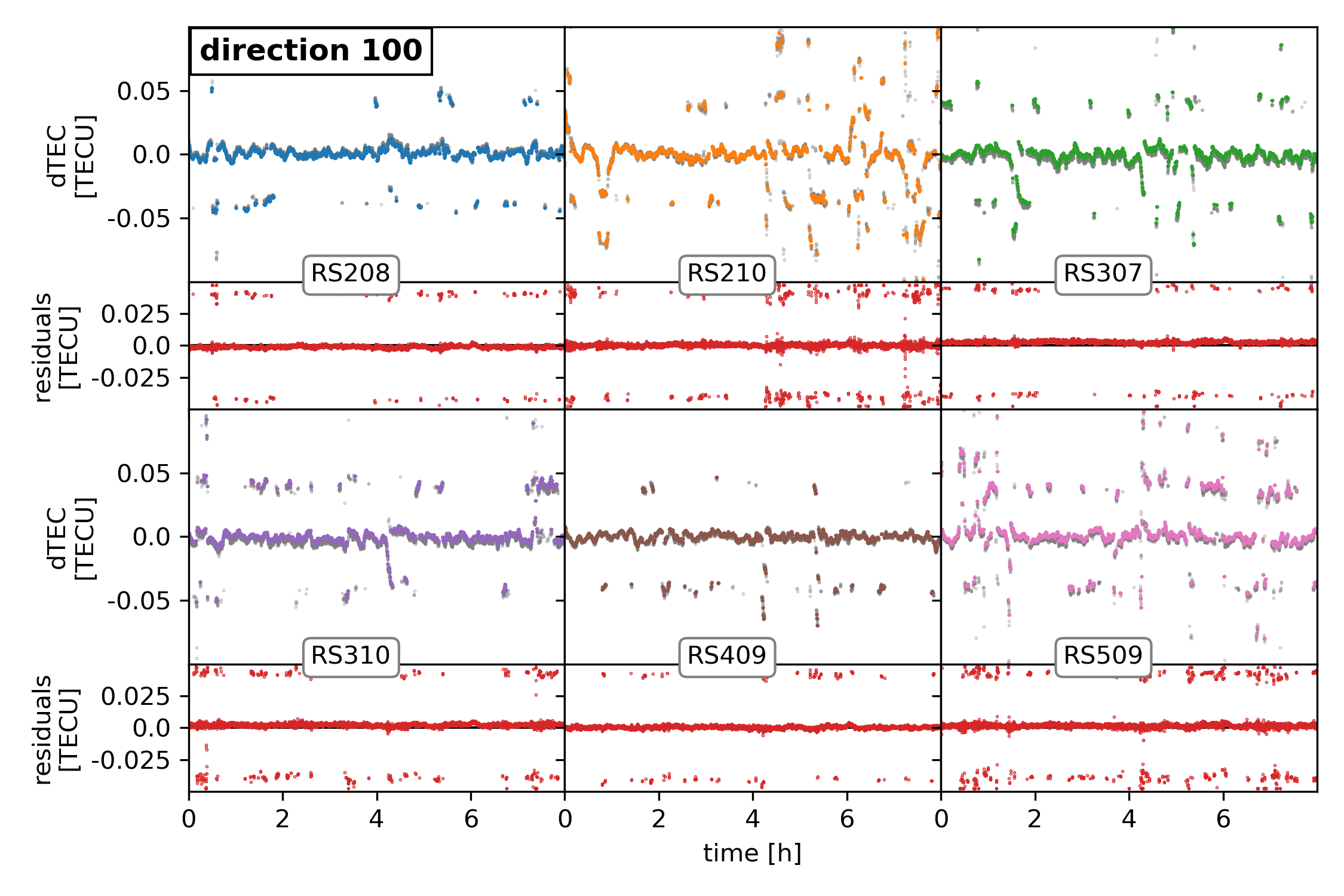}
   \caption{LBA - direction 100.}\label{b}
\end{subfigure}
\begin{subfigure}{0.495\linewidth}
   \includegraphics[width=1.0\linewidth]{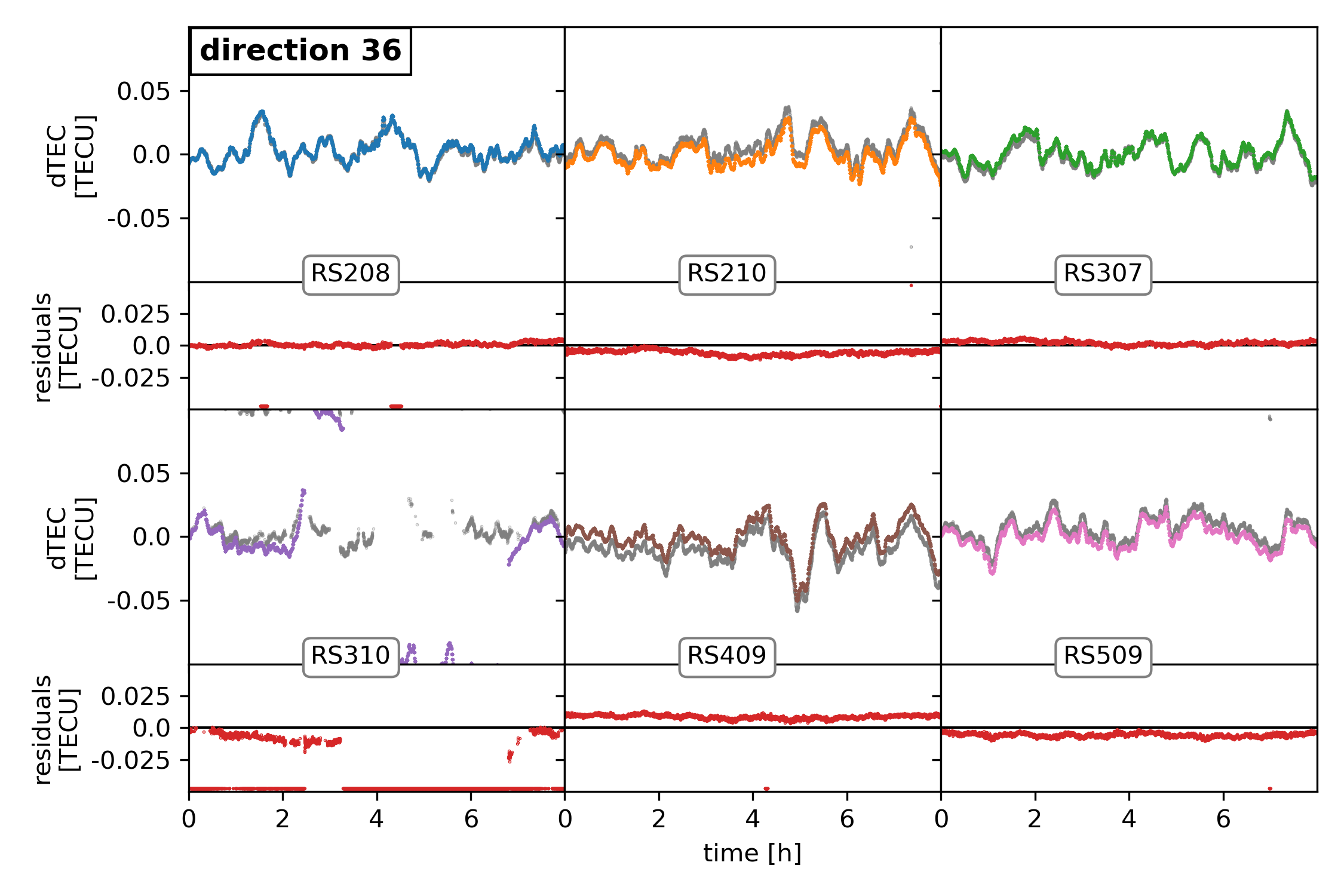}
   \caption{HBA - direction 36.}\label{c}
\end{subfigure}
\begin{subfigure}{0.495\linewidth}
   \includegraphics[width=1.0\linewidth]{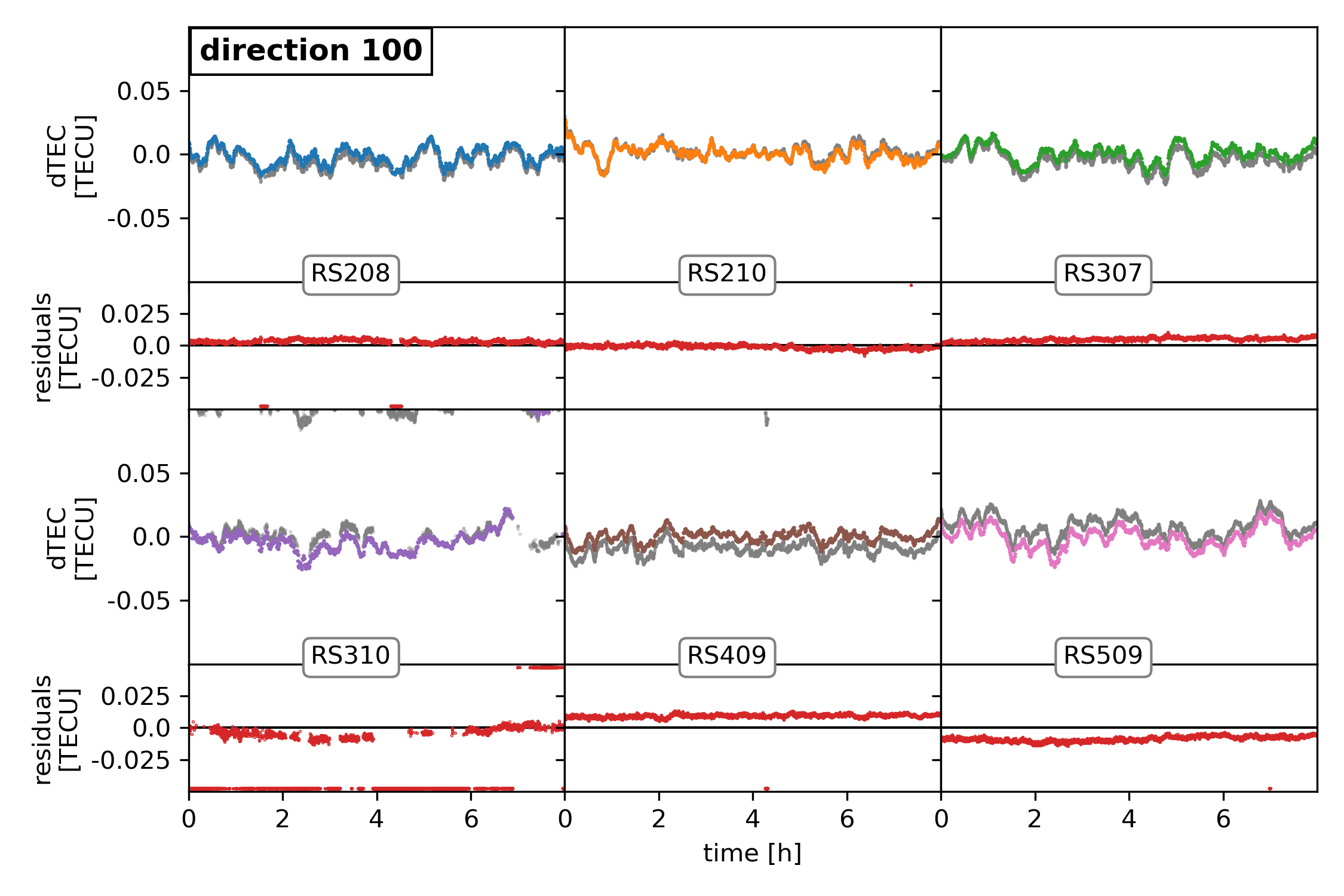}
   \caption{HBA - direction 100.}\label{d}
\end{subfigure}
\begin{subfigure}{0.495\linewidth}
   \includegraphics[width=1.0\linewidth]{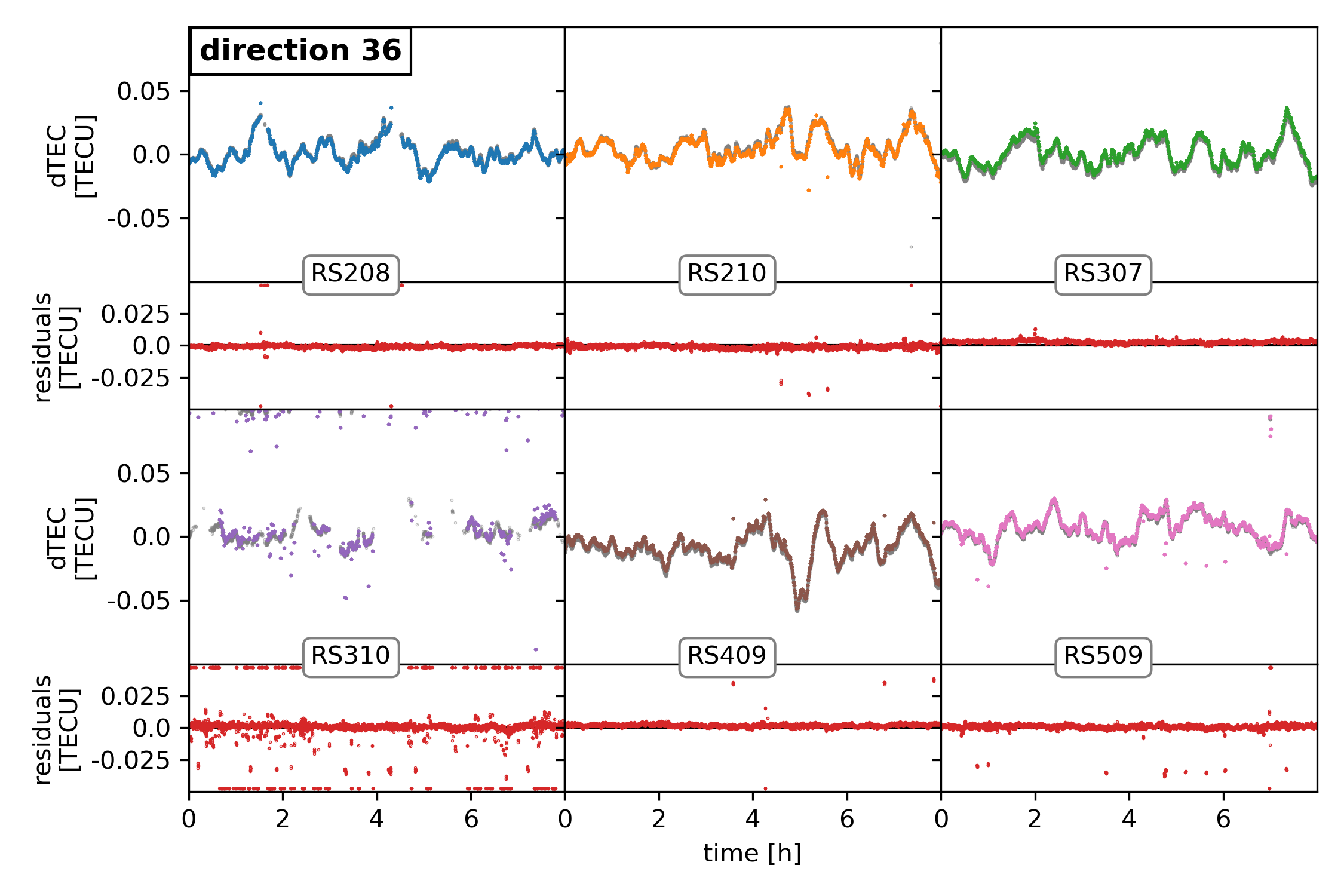}
   \caption{LBA + HBA joint calibration - direction 36.}\label{e}
\end{subfigure}
\begin{subfigure}{0.495\linewidth}
   \includegraphics[width=1.\linewidth]{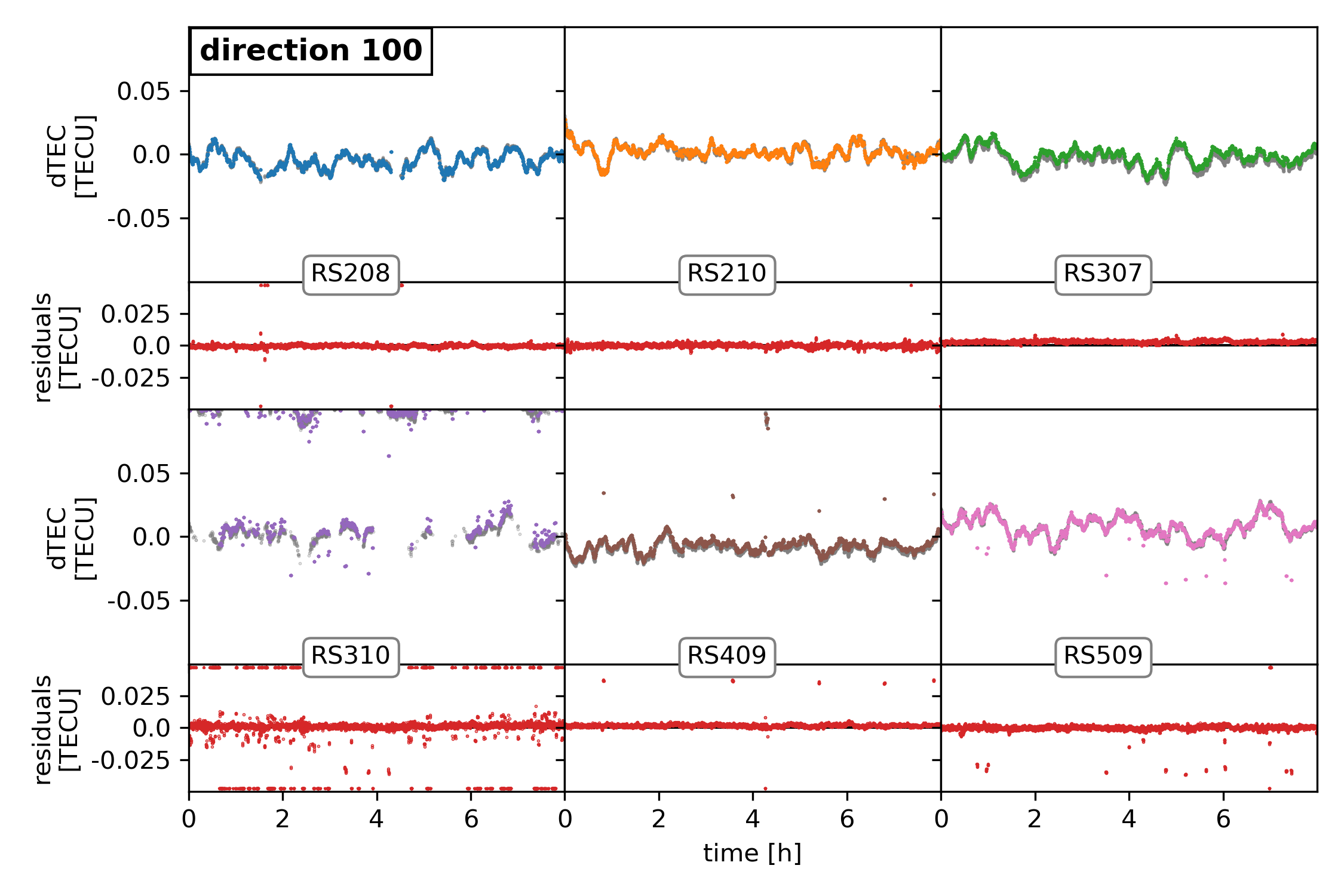}
   \caption{LBA + HBA joint calibration - direction 100.}\label{f}
\end{subfigure}
\caption{Each of the six panels (a) to (f) shows direction-dependent \TEC-solutions (colors except red) and residuals (red) with respect to the simulation input for six distant RS. Residuals outside of the displayed range are indicated by arrows. Figures (a), (c) and (e) in the left column show solutions towards \emph{dir36}, whereas figures (b), (d) and (f) in the right column show solutions towards \emph{dir100}. In the two figures in the top row, only LBA phase solutions were used to extract the \dTEC, while for the center row, only HBA was considered. The figures in the bottom row show joint calibration solutions derived from LBA and HBA combined. The gray lines in the background of the \dTEC panels show the difference between input \dTEC and direction-independent \dTEC solutions for this station and direction, these values were used to calculate the residuals as $\dTEC_{residual} = \dTEC_{DDE} - (\dTEC_{input} - \dTEC_{DIE})$. All values are referenced to \emph{CS001LBA} respectively \emph{CS001HBA0}.}\label{fig:ddetec}
\end{figure*}
\begin{figure*}
\centering
\begin{subfigure}{1.0\linewidth}
   \centering
   \includegraphics[width=0.85\linewidth]{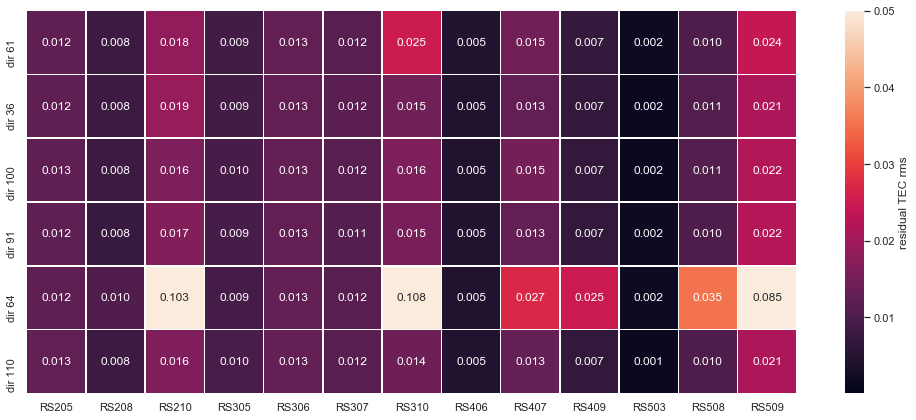}
   \caption{LBA \TEC-solutions}
\end{subfigure}
\begin{subfigure}{1.0\linewidth}
   \centering
   \includegraphics[width=0.85\linewidth]{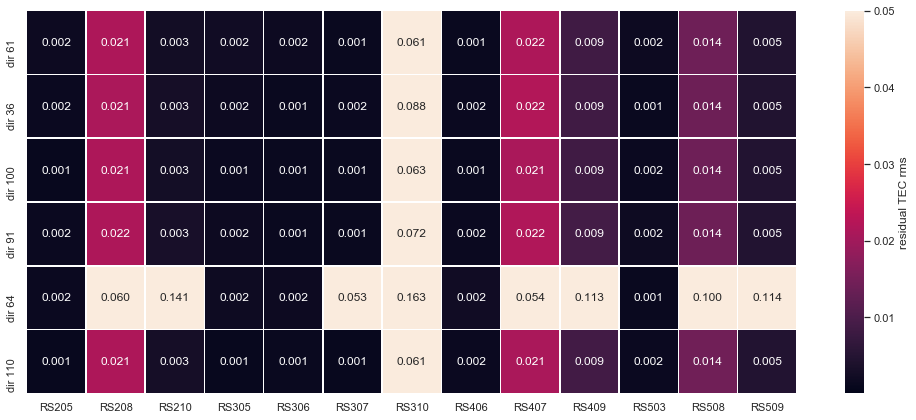}
   \caption{HBA \TEC-solutions}
\end{subfigure}
\begin{subfigure}{1.0\linewidth}
   \centering
   \includegraphics[width=0.85\linewidth]{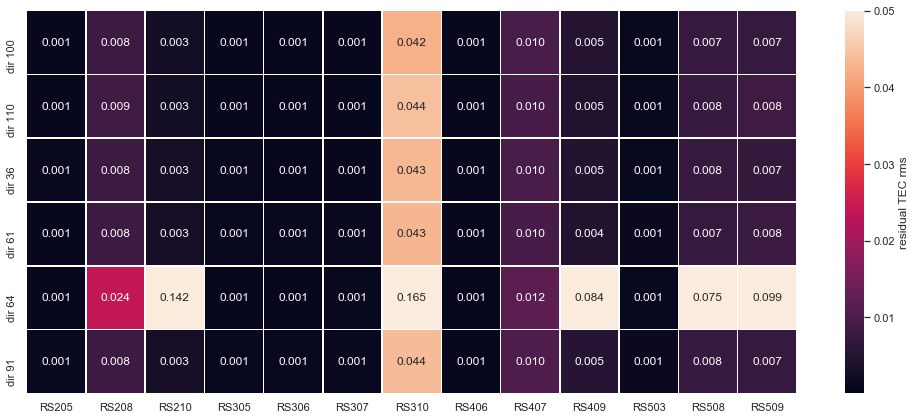}
   \caption{LBA + HBA joint calibration \TEC-solutions}
\end{subfigure}
\caption{Each of the three plots (a) - (c) shows the direction-dependent \TEC root-mean-square-error as a function of direction ($y$-axis) and station ($x$-axis). The \TEC was extracted from phase solutions derived from LBA (a), HBA (b) or both (c). }\label{fig:pivots}
\end{figure*}

\begin{figure*}
\centering
\begin{subfigure}{0.49\linewidth}
   \centering
   \includegraphics[width=0.8\linewidth]{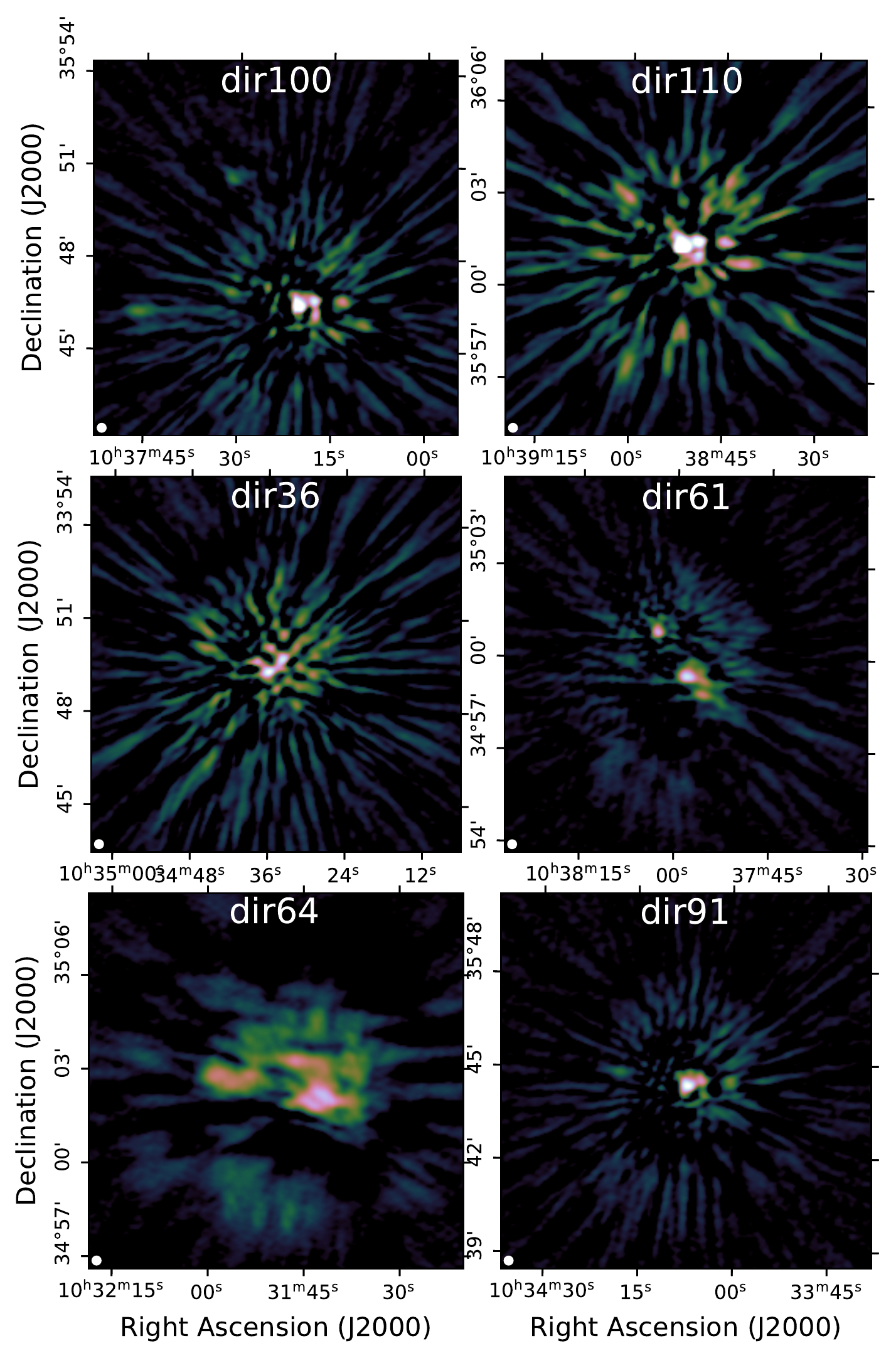}
   \caption{Direction-independent calibration only.}
\end{subfigure}
\begin{subfigure}{0.49\linewidth}
   \centering
   \includegraphics[width=0.8\linewidth]{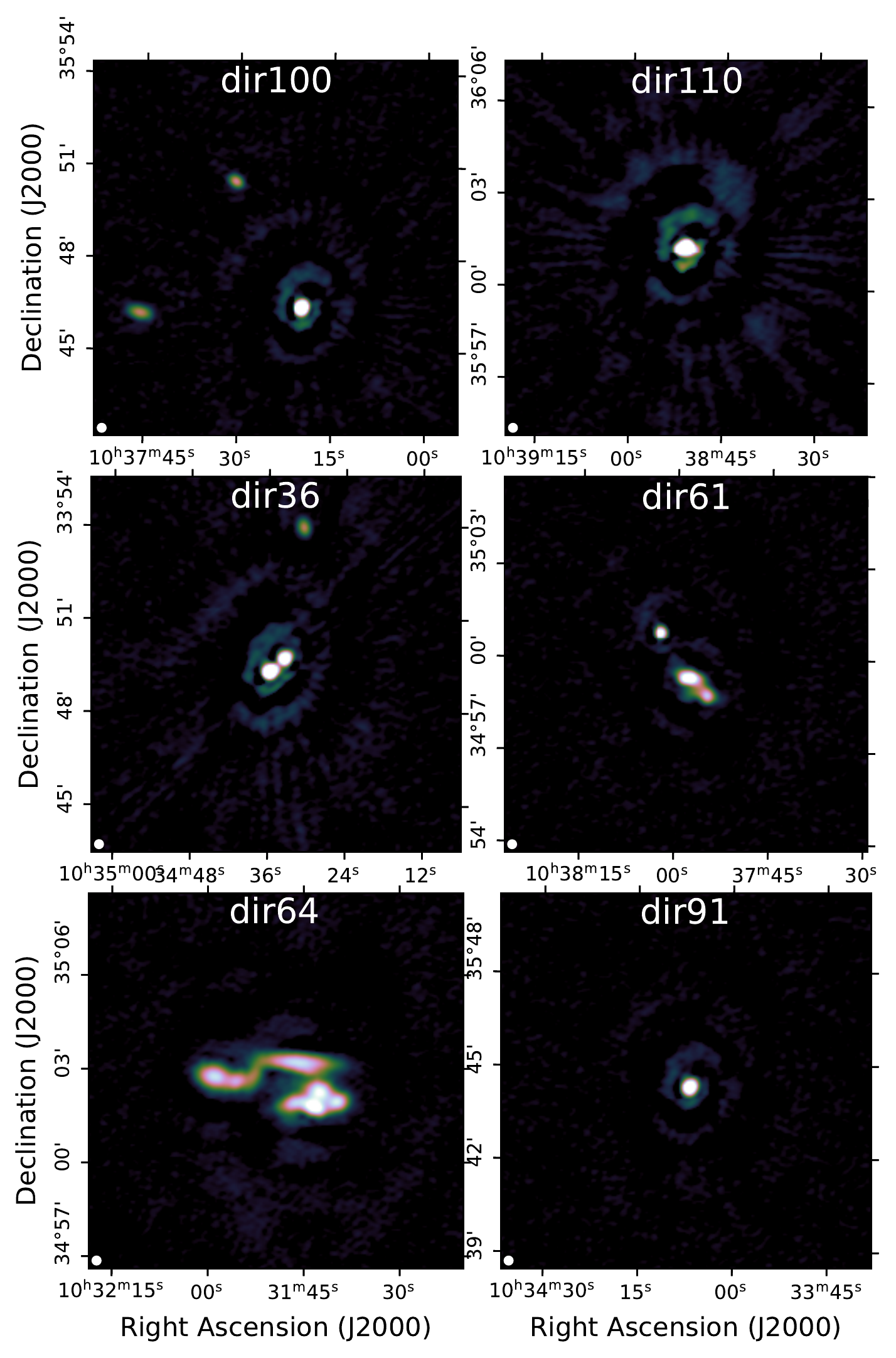}
   \caption{LBA \TEC-solutions}
\end{subfigure}
\begin{subfigure}{0.49\linewidth}
   \centering
   \includegraphics[width=0.8\linewidth]{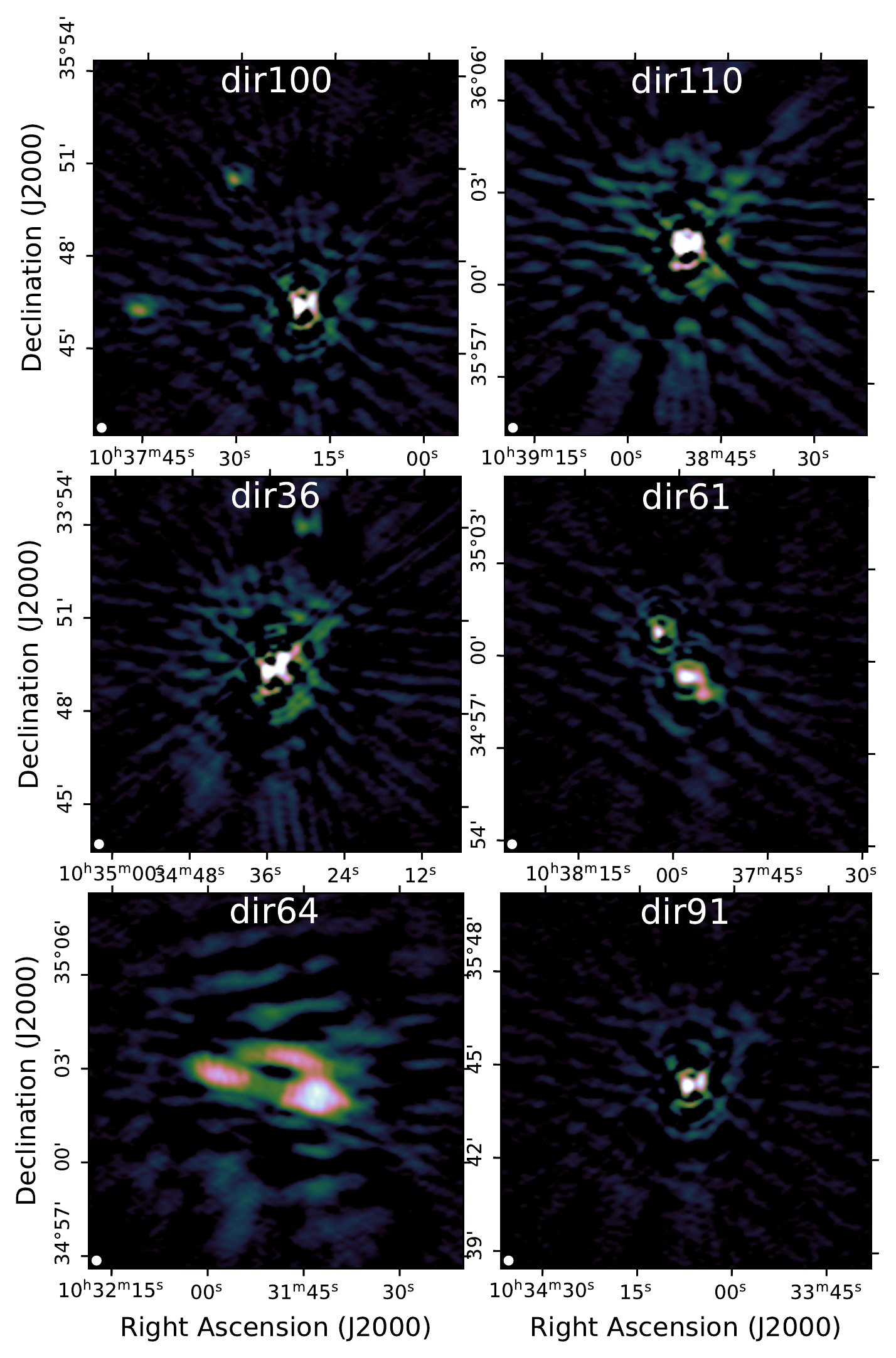}
   \caption{HBA \TEC-solutions transferred to LBA}
\end{subfigure}
\begin{subfigure}{0.49\linewidth}
   \centering
   \includegraphics[width=0.8\linewidth]{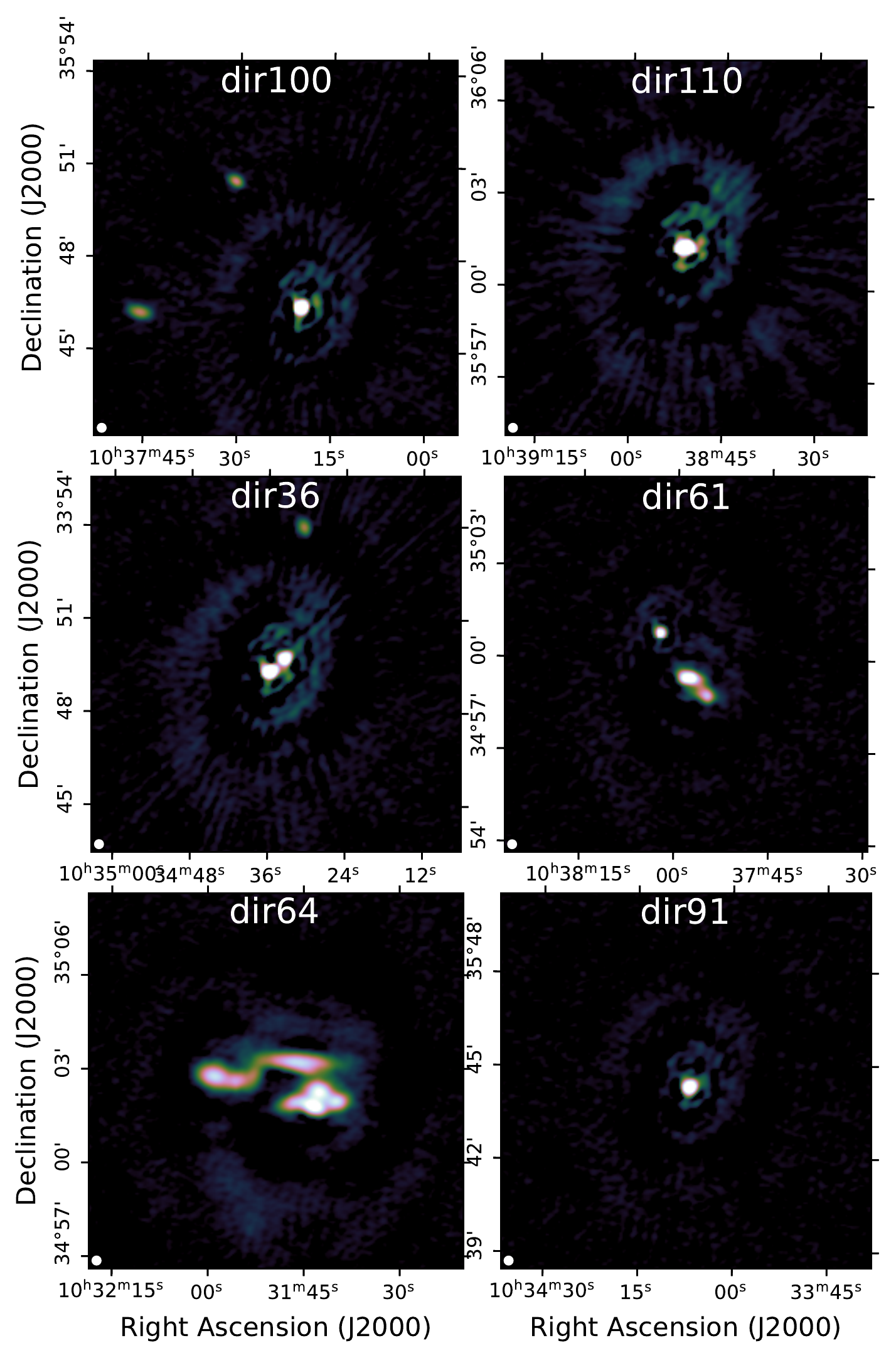}
   \caption{LBA + HBA joint calibration \TEC-solutions}
\end{subfigure}
\caption{The panels (a) - (d) show a 12''-region around the six calibrator directions used in the analysis with different calibration solutions applied. All images share the same color scale and show LBA data centered at \SI{54}{\mega\hertz}. Panel (a) is corrected only for direction-independent effects, panel (b) and (c) are corrected using \TEC-solutions derived from only LBA or only HBA, and panel (d) shows the joint-calibration case.}\label{fig:ddeimg}
\end{figure*}

\subsection{Calibration scenarios for LOFAR\,2.0}

To study possible scenarios of direction-dependent calibration in LOFAR\,2.0, we rely on strategies which extract the \TEC from the direction-dependent phase solutions. We use the software \texttt{LoSoTo} \citep{deGasperin2019} to fit the dispersive-delay term (see \autoref{eq:dispersive_delay}). To reduce the local minima problem, a grid-search is performed to determine the initial guess of the optimization. We assign a weight to the LBA phase solutions according to $w(\nu)=\mathrm{min}(\SEFD)/{\SEFD}(\nu)$, the HBA phase solutions are assigned a constant weight of $w = 1$. We compare three different approaches to extract \dTEC from the phase-solutions:
\begin{enumerate}
    \item Using only the LBA solutions, as shown in \autoref{a},b. 
    \item Using only the HBA solutions, as shown in \autoref{c},d. This scenario is the case of solution-transfer. 
    \item Testing joint calibration -  we combine LBA and HBA solutions to extract \TEC over a broad frequency range, see  \autoref{e},f.
\end{enumerate}
Due to the different beam sizes, the direction-independent \TEC-solution of LBA and HBA differ slightly. Therefore, the direction-dependent solutions will also be slightly different to account for this offset. Thus, direction-independent solutions obtained with one antenna type cannot directly be combined with direction-dependent solutions obtained with the other.
We take this into account by applying the HBA direction-independent solutions for the remote stations to the LBA data in the cases 2 and 3. For case 3, we also add the phases corresponding to the difference between the HBA and LBA direction-independent \TEC solutions to the direction-dependent phase solutions in LBA to establish a common ground for direction-dependent \TEC-extraction.
The results of all three approaches as well as the residuals with respect to the simulation input are displayed in \autoref{fig:ddetec} for the calibrators \textit{dir36} and \textit{dir100}. In \autoref{fig:pivots}, we show the \rms-error of the \TEC solutions for the six calibration directions and all remote stations. 
In the LBA only case, while many of the residuals are close to zero, a substantial fraction converged towards neighboring minima, leading to an \rms-error which is in most cases higher than in the competing approaches. This can be explained with the low signal-to-noise ratio of the LBA. For the HBA observation, most stations and directions reach a \si{\milli\tecu}-scale \rms-error but show a small systematic offset in the residuals. This offset appears to be present in most directions to a similar extent, pointing towards a direction-independent phase error in the data which is not fully corrected.  One negative exception is the complex source \textit{dir64}, where no sufficiently accurate model could be derived in HBA, leading to significantly noisier solutions. Further parameter fine-tuning could potentially improve solutions for this direction.
Furthermore, for station \textit{RS310}, improper solutions were found during direction-independent calibration. This issue propagates to direction-dependent calibration. In the joint-calibration approach, the mean \rms of the solutions is the lowest and the aforementioned issues are partially solved. The \TEC-jumps are considerably  less frequent than in the LBA-only case, the systematic drifts present in the HBA residuals are attenuated and the solutions for the problematic station RS310 are substantially improved compared to the HBA-only case.

In \autoref{fig:ddeimg}, we show LBA images of the six calibrators, comparing image quality without direction-dependent corrections to the three different direction-dependent calibration strategies of our analysis. In all cases, the image quality improved compared to the direction-independently calibrated image. The best image quality is achieved using either LBA or joint-calibration \TEC solutions, both approaches result in a noise level of \SI{0.9}{\milli\jansky\per\beam}. The characteristic star-like patterns around the sources caused by the ionospheric dispersive delay are removed almost completely. However, extended, spiral-shaped artifacts remain. The image quality using only the LBA stations for calibration is equally good and for some directions even slightly better than the joint calibration despite the joint-calibration solutions being more accurate. This can be explained by two points: first, the jumps in LBA-extracted \TEC lead to a high \rms of the residuals while still providing a decent calibration for part of the frequency band, and second, remaining corruptions in the LBA data might not be fully described by a \TEC-term.

Image quality is worse in the solution-transfer scenario: noticeable calibration artifacts are present, revealing that multiple stations are not well calibrated. 
This gives rise to the question of why the image quality in this scenario is inferior if the solutions show fewer jumps than the ones obtained from LBA. The explanation for this must lie in the systematic offsets which can be observed in the HBA residuals in \autoref{fig:ddetec}. A phase error in HBA that was not fully corrected by our calibration strategy is the likely cause of these offsets; however, the root of this issue could not be determined. While the HBA \TEC-solutions will certainly improve if the phase offset can be solved, this also highlights an intrinsic disadvantage of the solution-transfer approach: while the presence of small phase errors in HBA data can still lead to satisfying results using \TEC-calibration at HBA frequencies, transferring the solutions to LBA can strongly amplify any phase errors from \TEC-offsets. Even if there were no residual errors in the simulated data, minor systematic effects which are not present in the simulation could cause similar problems in real observations. Therefore, solution-transfer is only viable if the HBA data is free of non-ionospheric phase errors.

A more refined calibration strategy could improve the solutions obtained during direction-dependent calibration. Most notably, we emphasize that there are further points were the simultaneity of the observation could be exploited. It would be possible to jointly calibrate the first, second and possible third order ionospheric term as well as clock delays during direction-independent calibration. Additionally, the spectral properties of the model components could be estimated more accurately by a unified model for the low and high band, possibly obtained by joint de-convolution. 

To use joint calibration strategies in production in the future, the development of specialized software will be necessary. Possible advancements include the implementation of a solution algorithm which can solve for \TEC and further frequency-dependent effects on LBA and HBA data together, bypassing the intermediate phase-solution. Furthermore, joint calibration could enable the leap from facet-based towards \TEC-screen based calibration as proposed in \citet{Albert2020} in LOFAR\,2.0 by increasing the robustness of the \TEC-estimates towards residual phase errors.

\subsection{Limitations}

A number of points must be considered when evaluating the accuracy of \texttt{LoSiTo} simulations. First of all, only the first and second order ionospheric effects are implemented in the simulation. Higher order effects are non-negligible at the lowest frequencies observed by LOFAR ($\nu \lesssim \SI{40}{\mega\hertz}$). Additionally, in real LOFAR observations, the presence of ionospheric scintillation can affect the coherency of celestial radio signals under special ionospheric conditions. These scintillations, together with artificial radio-frequency interference (RFI), can render data unusable for periods of time, but are not accounted for in \texttt{LoSiTo}. Second, while we do not expect the thin-layer and frozen turbulence assumptions to interfere with facet-based calibration strategies, one needs to be cautious using the simulations when working with approaches that enforce spatial coherency across multiple stations, since the phase error from projecting the three-dimensional structure onto a two-dimensional layer is not represented.
Third, the simulated sky cannot recreate the complexity of the real radio sky due to limitations in computing power. The sky model used in this work underestimates the number density especially of faint sources. Furthermore, it does not contain emission in  side lobes or on very large angular scales, both of which are known to interfere with calibration \citep{deGasperin2020, shimwell2019}.  Fourth, the beam model employed in the simulation is the product of semi-analytic simulations. It is known that the real beam response deviates from this model to some extent \citep{deGasperin2019}. This deviation is not included in the simulation. 
Last, real LOFAR data can contain subdominant systematic effects which are not well understood at present and hence, cannot be modeled in simulations.

\section{Conclusions}

In this paper, we presented models for a comprehensive list of systematic effects in LOFAR and LOFAR\,2.0 observations and  the \texttt{LoSiTo} code in which we embedded them. These models include a turbulent thin-layer representation of the ionosphere which is used to derive the first and second order ionospheric effect. Furthermore, \texttt{LoSiTo} features the systematic effects of clock error, polarization misalignment, the primary beam and bandpass responses as well as an estimate of the LOFAR\,2.0 noise level based on empirically determined values for the LOFAR \SEFD. The product of a simulation is a \textit{measurement set}, which can be further processed with standard radio astronomy tools. The code was developed with the aim to assist the progression of current and future LOFAR calibration strategies and is made publicly available. 

We used \texttt{LoSiTo} to simulate a full 8-h calibrator and target field observation using the LOFAR\,2.0 system. We presented the analysis of the simulated data, where we performed data reduction of the calibrator observation and direction-dependent calibration of the target field using adjusted LOFAR calibration pipelines. 
As a proof-of-concept, we investigated new strategies for direction-dependent calibration of the data. We compared ionospheric solutions derived from LBA and HBA separately to solutions derived jointly from both systems. We found that ionospheric parameters of the simulation can be determined most accurately in the joint calibration approach, where we reach a \si{\milli\TEC}-scale \rms error in 90\% of the cases. When we use only LBA data for calibration, the solutions are more noisy; nevertheless, the resulting image quality is very similar to the joint calibration approach with an \rms noise of \SI{0.9}{\milli\jansky\per\beam} away from bright sources and artifacts in the vicinity of the calibrators. This indicates that, while we managed to determine the \TEC accurately, our image-space results are still limited by the presence of systematic errors which could be resolved by an improvement of the strategy. 
For the case of solution transfer, where the ionospheric solutions found in the HBA calibration are applied to the LBA data, we find good convergence and very little noise in the solutions. However, they show systematic offsets at a scale of $\approx \SI{5}{\milli\tecu}$ which create strong artifacts in image-space. While further development of the calibration procedure could improve the image quality, this result reveals a central downside of the solution-transfer approach: errors in the HBA data are strongly amplified when the solutions are applied to LBA data. Therefore, solution transfer can only be an option if all non-ionospheric effects in HBA are corrected to high accuracy.

\section{Acknowledgements}
 We thank the anonymous referee and the editor for their constructive remarks on this work. This project is funded by the Deutsche Forschungsgemeinschaft (DFG, German Research Foundation) under projet number 427771150. The authors thank J. Hessels, C. Bassa, T. Shimwell and the LOFAR 2.0 developing team for their help.


\bibliographystyle{aa}
\bibliography{main}

\end{document}